\documentclass[10pt,letterpaper,twocolumn]{revtex4}
\pdfoutput=1 
\usepackage[latin1]{inputenc}
\usepackage{amsmath}
\usepackage{amsfonts}
\usepackage{amssymb}
\usepackage{graphicx}
\usepackage{amsthm}
\usepackage[ddmmyyyy,hhmmss]{datetime}

\usepackage{xcolor}

\newcommand{\ha}{\frac{1}{2}}
\newcommand{\vx}{\vec x}
\newcommand{\vy}{\vec y}
\newcommand{\vk}{\vec k}
\newcommand{\vq}{\vec q}

\newcommand{\R}{\mathbb{R}}

\newcommand{\ev}[1]{\langle#1\rangle}
\newcommand{\bra}[1]{\langle#1|}
\newcommand{\ket}[1]{|#1\rangle}

\newcommand{\mc}[1]{\mathcal{#1}}

\newcommand{\gauge}{\mbox{\tiny U(1)}}
\newcommand{\CFT}{\mbox{\tiny CFT}}
\newcommand{\UV}{\mbox{\tiny UV}}
\newcommand{\LD}{\mbox{\tiny LD}}

\begin{document}
	\title{Entanglement renormalization for gauge invariant quantum fields}
	\author{Adri\'an Franco-Rubio}
\affiliation{Perimeter Institute for Theoretical Physics,
	Waterloo, ON, N2L 2Y5, Canada}
\affiliation{Department of Physics and Astronomy, University of Waterloo, Waterloo, ON, N2L 3G1, Canada}
	\author{Guifr\'e Vidal}
\affiliation{Perimeter Institute for Theoretical Physics, Waterloo, ON, N2L 2Y5, Canada}
\affiliation{X, The Moonshot Factory, Mountain View, CA, 94043, USA}
\begin{abstract}
The continuous multiscale entaglement renormalization ansatz (cMERA) [Haegeman et al., Phys. Rev. Lett., 110, 100402 (2013)] is a variational wavefunctional for ground states of quantum field theories. So far, only scalar bosons and fermions have been considered. In this paper we explain how to generalize the cMERA framework to \textit{gauge invariant} quantum fields. The fundamental difficulty to be addressed is how to make the gauge constraints (\textit{local} linear constraints in the Hilbert space) compatible with the UV structure of the cMERA wavefunctional (which is generated by a \textit{quasi-local} entangler). For simplicity, we consider U(1) gauge theory in $d+1$ spacetime dimensions, a non-interacting theory with \textit{massless} Hamiltonian $H_{\gauge}$ and Gaussian scale invariant ground state $\ket{\Psi_{\gauge}}$. We propose a gauge invariant cMERA wavefunctional $\ket{\Psi^{\Lambda}_{\gauge}}$ that, by construction, accurately reproduces the long distance properties of $\ket{\Psi_{\gauge}}$ while remaining somewhat unentangled at short distances. Moreover, $\ket{\Psi^{\Lambda}_{\gauge}}$ is the exact ground state of a gauge invariant, local Hamiltonian $H^{\Lambda}_{\gauge}$ whose low energy properties coincide with those of $H_{\gauge}$. Our construction also extends the cMERA formalism to \textit{massive} (non-gauge invariant) vector boson quantum fields.

	\end{abstract}
	\maketitle
\section{Introduction}
% Gauge theories

Gauge theories stand among the most successful theories of physical reality, describing a wide range of phenomena -- from the standard model of particle physics \cite{weinberg,schwartz} and general relativity \cite{misner} to topological phases of quantum matter \cite{wen}. They are characterized by an explicit redundancy in the choice of degrees of freedom used to represent the physical system. This redundancy is the price to be paid in order to retain a more tractable and intuitive description, for instance one in terms of a local Hamiltonian. 
Gauge theories fit into the more general framework of \textit{constrained theories}, whose quantization is itself a rich and interesting subject \cite{gaugeBook}. A quantum gauge theory can be formulated so that physical states are confined to a particular, \textit{gauge invariant} subspace of the total Hilbert space of the theory. 

To go beyond perturbative treatments of gauge theory, one must often resort to numerical simulations. In lattice gauge theory \cite{Wilson, KogutSusskind}, spacetime is discretized into a lattice in such a way that gauge invariance is preserved. Then stochastic methods, such as Monte Carlo sampling, are used to study certain aspects of the discretized theory. For instance, and most prominently, such techniques have been used to successfully extract the mass spectrum of quantum chromodynamics (QCD) \cite{QCD1, QCD2}. In spite of their remarkable success, simulation strategies based on stochastic sampling suffer from the fermionic sign and complex action problems at finite fermionic density \cite{sign1, sign2} and, more generally, are not capable of simulating dynamics. For such important problems, alternative formulations are still much needed.

% Tensor networks 

In the past two decades, tensor networks have arisen as a useful new framework to treat quantum many-body problems on the lattice. By exploiting the entanglement structure of certain many-body wavefunctions, such as ground states and low energy states of local Hamiltonians, tensor networks offer efficient parameterizations and solutions to problems of otherwise unmanageable computational complexity. Much work has been devoted to applying tensor network algorithms to lattice gauge theories \cite{LGTTN1Ham1, LGTTN1Ham2, LGTTN1Ham3, LGTTN1Ham4, LGTTN1Ham5, LGTTN1Ham6, LGTTN1Ham7, LGTTN1Ham8, LGTTN1Ham9, LGTTN1Ham10, LGTTN1Ham11, LGTTN1Ham12, LGTTN1Ham13, LGTTN1Ham14, LGTTN1Ham15, LGTTN1Ham16, LGTTN1Ham17, LGTTN1Ham18, LGTTN1Ham19, LGTTN1Ham20, LGTTN1Ham21, LGTTN1Ham22, LGTTN1Ham23, LGTTN1Ham24, LGTTN1Ham25, LGTTN1Ham26, LGTTN1Ham28, LGTTN1Ham27, LGTTN1Lag1, LGTTN1Lag2, LGTTN1Lag3, LGTTN1Lag4, LGTTN1Lag5, LGTTN1Lag6, LGTTN1Lag7, LGTTN1Lag8, LGTTN1Lag9, LGTTN1Lag10, LGTTN1Lag11, LGTTN2Ham1, LGTTN2Ham2, LGTTN2Ham3, LGTTN2Ham4, LGTTN2Ham5, LGTTN2Ham6, LGTTN2Ham7, LGTTN2Ham8, LGTTN2Ham9, LGTTN2Ham10, LGTTN2Ham11,  LGTTN2Lag, LGTTNrev1, LGTTNrev2, LGTTNrev3, LGTTNrev4, LGTTNrev5}, with the expectation of advancing our numerical capabilities past the breaking points of standard techniques, such as the sign problem mentioned above in the case of Monte Carlo simulation (see \cite{LGTTNrev4, LGTTNrev5} for recent reviews). Successful simulations in one spatial dimension \cite{LGTTN1Ham1, LGTTN1Ham2, LGTTN1Ham3, LGTTN1Ham4, LGTTN1Ham5, LGTTN1Ham6, LGTTN1Ham7, LGTTN1Ham8, LGTTN1Ham9, LGTTN1Ham10, LGTTN1Ham11, LGTTN1Ham12, LGTTN1Ham13, LGTTN1Ham14, LGTTN1Ham15, LGTTN1Ham16, LGTTN1Ham17, LGTTN1Ham18, LGTTN1Ham19, LGTTN1Ham20, LGTTN1Ham21, LGTTN1Ham22, LGTTN1Ham23, LGTTN1Ham24, LGTTN1Ham25, LGTTN1Ham26, LGTTN1Ham27, LGTTN1Ham28, LGTTN1Lag1, LGTTN1Lag2, LGTTN1Lag3, LGTTN1Lag4, LGTTN1Lag5, LGTTN1Lag6, LGTTN1Lag7, LGTTN1Lag8, LGTTN1Lag9, LGTTN1Lag10, LGTTN1Lag11} and partial success in two spatial dimensions \cite{LGTTN2Ham1, LGTTN2Ham2, LGTTN2Ham3, LGTTN2Ham4, LGTTN2Ham5, LGTTN2Ham6, LGTTN2Ham7, LGTTN2Ham8, LGTTN2Ham9, LGTTN2Ham10, LGTTN2Ham11,  LGTTN2Lag} are certainly encouraging. However, very significant improvements will be required before e.g. QCD in three spatial dimensions can be meaningfully tackled. Finally, \textit{continuous} tensor networks have been introduced more recently to simulate quantum field theories (QFTs) directly in the continuum, that is, without introducing a lattice \cite{cMPS1,cMERA}. Compared to lattice tensor networks, continuous tensor network techniques are still in their infancy, but one might hope that once they are better understood in one spatial dimension they will be more easily extended to higher dimensions.

% MERA

The particular tensor network we will be interested in here is the multi-scale entanglement renormalization ansatz (MERA) \cite{MERA1,MERA2}. On the lattice, the MERA can be interpreted as a quantum circuit that builds an entangled many-body wavefunction starting from an unentangled or product state by introducing entanglement via (nearest-neighbor) local unitary gates that effectively act at progressively smaller length scales. Running this procedure backwards, we obtain a dual interpretation of the MERA as encoding a (discrete version of a) renormalization group flow, where entanglement at short length scales is progressively removed from the wavefunction. For lattice gauge theories in the Hamiltonian formalism, MERA has been seen to offer a proper framework to represent gauge invariant ground states \cite{LGTTN2Ham1, LGTTN2Ham2}. In this case the renormalization group transformations exactly preserve the gauge constraints along the flow. 

% cMERA

More specifically, in this work we will be concerned with the \textit{continuous} MERA (cMERA), which was proposed by Haegeman, Osborne, Verschelde and Verstraete in Ref. \cite{cMERA}. While the original formulation of Ref. \cite{cMERA} applies to both interacting and non-interacting fields, in practice cMERA is only well-understood for non-interacting QFTs, in which case it this variational ansatz is a Gaussian wavefunctional (see however \cite{intercMERA1,intercMERA2,intercMERA3} for proposals that go beyond a purely Gaussian wavefunctional, e.g. through the use perturbation theory). It is of course not obvious why one would need a variational ansatz for a free QFT, which can be solved exactly without much effort. Nonetheless, the Gaussian cMERA offers a useful demonstration that MERA can be brought into the continuum, while also providing insights into the entanglement structure of ground state wavefunctionals. Moreover, Gaussian cMERA has found application as a conjectured realization of the holographic principle of quantum gravity, namely as a toy model for the Anti-de-Sitter / conformal field theory (Ads/CFT) correspondence \cite{cMERAholo1, cMERAholo2, cMERAholo3, cMERAholo4, cMERAholo5, cMERAholo6, cMERAholo7, cMERAholo8, cMERAholo9}. Quite promisingly, Ref. \cite{Zou2019} recently proposed a particular realization of Gaussian cMERA, dubbed magic cMERA, which has a UV structure analogous to that of the continuous matrix product state (cMPS), another continuous tensor network. This is important because cMPS techniques work equally well for both non-interacting and interacting QFTs. Thanks to this connection it is now finally possible to produce and efficiently manipulate strongly-correlated (that is, highly non-Gaussian) cMERA wavefunctionals using cMPS techniques \cite{Ganahl2019}.

Ref. \cite{cMERA} formulated cMERA for scalar bosons and for fermions. In this paper we take a step further and extend the cMERA formalism to gauge invariant quantum fields. Our main motivation is simple. If, as we expect, the cMERA program is to eventually give rise to a useful numerical simulation framework for interacting QFTs, then understanding how to handle gauge invariant quantum fields is a priority, given the central role gauge theories play in modern physics. A second motivation for our work comes from current applications of cMERA as a toy models for the AdS/CFT correspondence. There the CFT theory is often taken to be a gauge theory with a large gauge group. Therefore a gauge invariant cMERA could also be useful to build improved toy model of the AdS/CFT correspondence.

% U(1)

In this work we will illustrate how the cMERA formalism can be extended to gauge theories by considering the simple case of noninteracting $U(1)$ gauge theory, or electromagnetism without matter fields, in $d+1$ spacetime dimensions as a proof-of-principle example. $U(1)$ gauge theory is ideal for our purposes, because the Hamiltonian is quadratic and this allows us to show, explicitly and exactly, how the \textit{local} linear constraints in Hilbert space implementing gauge invariance can coexist with the \textit{quasi-local} character of the \textit{entangler} that generates the cMERA wavefunctional. Our eventual goal is to address interacting gauge theories, where the interaction may be due to either coupling to matter fields or to considering non-Abelian gauge groups (or to both at the same time). While a cMERA framework for interacting gauge theories may need to build on the proposals of Ref. \cite{Ganahl2019}, we expect that the compatibility between the quasi-local character of the cMERA and the local character of the gauge constraints, as demonstrated here for a non-interacting theory, will work in a similar way in the interacting case.

We devote Section \ref{sec2} to reviewing, and setting our notation for, both cMERA and U(1) gauge theory. In Section \ref{sec3} we introduce a Gaussian cMERA that approximates the ground state of U(1) gauge theory, and elaborate on its properties. In particular, we will see that our proposal is a natural extension to gauge fields of the \textit{magic} entanglement renormalization scheme of Ref. \cite{Zou2019}, thus paving the way to subsequently building strongly-correlated cMERA wavefunctionals for interacting gauge theories. Finally, we include a series of appendices elaborating on particular aspects of our construction.

\section{Review of background material}
\label{sec2}

In this section we briefly review some required background material. First we introduce the cMERA formalism of Ref. \cite{cMERA} and a particular realization thereof, the magic cMERA \cite{Zou2019}. Then we review U(1) gauge theory, which is \textit{massless}, and the closely related \textit{massive} vector boson QFT.

\subsection {cMERA}

The cMERA \cite{cMERA} (denoted $\ket{\Psi^{\Lambda}(s)}$ or $\ket{\Psi^{\Lambda}}$ depending on the context) is an ansatz that aims to approximate some target wavefunctional of interest, typically the ground state of a QFT Hamiltonian in $d$ spatial dimensions. It is produced through a unitary \textit{entangling evolution in scale}, which yields a one-parameter family of cMERA states
\begin{equation}
\ket{\Psi^\Lambda(s)} \equiv  \mc P\exp{\left( -i\int_{0}^{s}{ds'\;\left[L+K(s')\right]}\right) }\ket\Lambda.
\label{eq:evol}
\end{equation}
Here $\mc P\exp$ denotes a path-ordered exponential, $s$ is the scale parameter, $L+K(s)$ is the Hermitian operator that generates the entangling evolution and $\ket{\Lambda}$ is the initial state. This initial state is taken to be unentangled, with correlation functions that vanish when evaluated at different points:
\begin{equation}
\bra\Lambda\mc O(x)\mc O(y)\ket\Lambda=0,\qquad x\neq y.
\end{equation}
(We can think of $\ket{\Lambda}$ as the continuum limit of an unentangled/product state of a lattice system.)

 \par The generator of the evolution in scale is split into two contributions, $L$ and $K(s)$, which play two distinct roles. $L$ is the generator of \textit{non-relativistic} scale transformations that rescale both space and the field degrees of freedom. As such, $L$ only depends on the field content of the theory under study, and not on the specific form of the Hamiltonian whose ground state we aim to approximate with the cMERA. For a generic field $\varphi$ with non-relativistic scaling dimension $\Delta_{\varphi}$, we have
 \begin{equation}
 e^{isL}\varphi(\vx)e^{-isL}=e^{s\Delta_{\varphi}}\varphi(e^s\vx).
 \end{equation}
 On the other hand $K(s)$ is a quasilocal operator called \textit{entangler} (see Eqs. \eqref{eq:entangler} and \eqref{eq:magicg} below for an example in one spatial dimension). By \textit{quasilocal} we mean that the operator, an integral of a quasi-local density, acts at a specific length scale. It is standard to denote this length scale as $\Lambda^{ -1}$, where $\Lambda$ is the corresponding momentum scale. Intuitively, the entangling evolution in scale builds the cMERA wavefunctional from the initial uncorrelated state $\ket{\Lambda}$ by progressively introducing entanglement at length scale $\Lambda^{-1}$ as we keep ``zooming in'', both rescaling space and the fields. The resulting state will contain correlations at a range of length scales above $\Lambda^{-1}$, but will (partially) preserve the unentangled character of the initial state $\ket{\Lambda}$ at shorter distances. This idea can be made more precise, and allows us to say that a cMERA state presents an entanglement UV cutoff at length scale $\Lambda^{-1}$ \cite{Adrian}.

Let $\ket{\Psi}$ be the target state that the ansatz wavefunctional $\ket{\Psi^{\Lambda}(s)}$ aims to approximate at distances $x$ larger than $\Lambda^{-1}$, in the sense that e.g. the correlators agree to high accuracy for $\Lambda x \gg 1$. When the ansatz succeeds, we say that $\ket{\Psi^{\Lambda}(s)}$ is a good long distance (LD) approximation to $\ket{\Psi}$ and denote it by
\begin{equation}
 \ket{\Psi^{\Lambda}(s)}~~ \stackrel{\LD}{\sim} ~~\ket{\Psi}.
\end{equation}

\par
\begin{figure}[htp]
	\centering
	\includegraphics[width=4cm]{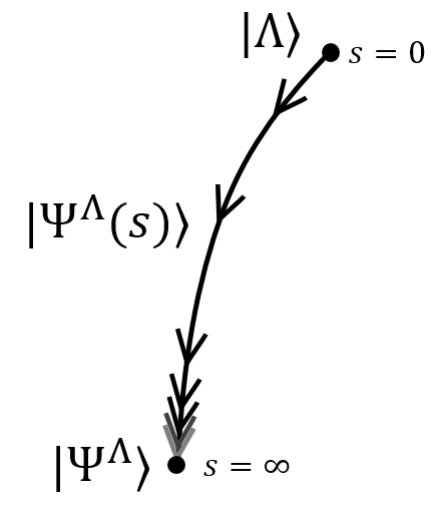}
	\caption{The entangling evolution in scale generates a cMERA state $\ket{\Psi^{\Lambda}(s)}$ for each value $s\in[0,\infty)$. When the entangler is scale independent, $K(s)=K$, then in the large $s$ limit we obtain a fixed point, scale invariant cMERA $\ket{\Psi^{\Lambda}}$. }
	\label{fig:cMERAflow}
\end{figure}

 Of particular interest in this paper is the case where the entangler is independent of $s$, $K(s)\equiv  K$. Then, in the limit $\lim_{s\rightarrow \infty} \ket{\Psi^{\Lambda}(s)}$ (assuming this limit exists) we obtain a fixed point wavefunctional known as a \textit{scale invariant} cMERA $\ket{\Psi^{\Lambda}}$,
\begin{equation}
\ket{\Psi^\Lambda} \equiv  \lim_{s\to\infty}{e^{ -is(L+K)} }\ket\Lambda,
\label{eq:asympt}
\end{equation}
see Fig. \ref{fig:cMERAflow}. By construction, the fixed point wavefunctional $\ket{\Psi^{\Lambda}}$ is invariant under further evolution under $L+K$. We can think of $D^{\Lambda} \equiv  L+K$ as a generator of an alternative notion of scale transformations, one that is adapted to the specific theory under study. For instance, in the context of a free boson CFT \cite{Qi}, $D^{\Lambda}$ is the generator of (a quasi-local version of) relativistic scale transformations, whereas $L$ generates non-relativistic scale transformations. Then we say that $\ket{\Psi^{\Lambda}}$ is invariant under this alternative notion of scale transformations, which is why we call $\ket{\Psi^{\Lambda}}$ a scale invariant cMERA wavefunctional. 

\par Ideally, given a target QFT Hamiltonian, the specific form of the entangler $K(s)$ should be determined variationally from a procedure such as energy minimization. While no general algorithms have been developed so far to determine the entangler variationally, in the particular case of free fields one can find examples of entanglers that give rise to interesting cMERA wavefunctionals. For example, for a free scalar $\phi(x)$ in one spatial dimension, with conjugate momentum $\pi(x)$, a possible family of entanglers is given by the translationally invariant, quasi-local quadratic operator \cite{cMERA}:
\begin{equation}
K(s)=\dfrac{-i}{2}\int{dx\,dy\;g(x-y,s)\psi(x)\psi(y)+\text{h.c}}
\label{eq:entangler}
\end{equation}
where $\psi(x)$ is the annihilation operator
\begin{equation}
\psi(x) \equiv  \sqrt{\dfrac{\Lambda}{2}}\phi(x)+i\sqrt{\dfrac{1}{2\Lambda}}\pi(x)
\end{equation}
and $g(x,s)$ is some profile function at scale $s$. A simple example, independent of $s$, is the profile $g(x,s) = g(x) \sim e^{-(\Lambda x)^2}$ used in Ref. \cite{cMERA}. In Eq. \eqref{eq:magicg} below we introduce an alternative choice of profile.

\subsection{Magic cMERA}

The variational parameters in cMERA correspond to different choices of the entangler $K$. There is a lot of freedom in choosing the entangler $K$, compatible with obtaining a good long distance approximation to a fixed target state, and one may be able to use this freedom to identify particularly useful subclasses of entanglers. 

In one spatial dimension, consider a relativistic free \textit{massless} boson CFT, with Hamiltonian
\begin{equation}
H^{\CFT} \equiv  \frac{1}{2}\int dx~\left[ \pi(x)^2+(\partial_x\phi(x))^2\right],
\label{eq:HCFT}
\end{equation}
and ground state $\ket{\Psi^{\CFT}}$, or more generally a relativistic free \textit{massive} boson QFT, with Hamiltonian 
\begin{equation}
H_m \equiv  H^{\CFT} + \frac{m^2}{2}\int dx~\phi(x)^2,
\label{eq:Hm}
\end{equation}
and ground state $\ket{\Psi_m}$. It was recently shown \cite{Zou2019} that the choice of entangler profile
\begin{equation}
g(x,s)\equiv  g(x)=\dfrac{\Lambda}{4}e^{-\Lambda|x|}
\label{eq:magicg}
\end{equation}
leads to cMERA wavefunctionals $\ket{\Psi^{\Lambda}}$ and $\ket{\Psi^{\Lambda}(s)}$ with
\begin{equation}
  \ket{\Psi^{\Lambda}}~~ \stackrel{\LD}{\sim} ~~\ket{\Psi^{\CFT}},
 \label{eq:PsiLamSim} 
\end{equation}
for the massless case and
\begin{equation}
  \ket{\Psi^{\Lambda}(s)}~~ \stackrel{\LD}{\sim} ~~\ket{\Psi_{m(s)}},
\label{eq:PsiLamSim2}
\end{equation}
where $\ket{\Psi_{m(s)}}$ is the relativistic massive ground state $\ket{\Psi_m}$ for mass
\begin{equation}
m(s) \equiv  \Lambda e^{-s}.
\label{eq:mass}
\end{equation}
Moreover, the magic cMERA wavefunctional has two remarkable properties, that we summarize next.

(i) \textit{Compatibility with cMPS:} $\ket{\Psi^{\Lambda}(s)}$ has the same UV structure as a \textit{continuous matrix product state} (cMPS) \cite{Zou2019}. As a result, cMPS techniques \cite{cMPS1, cMPS2, cMPS3, cMPS4, cMPS5, cMPS6, cMPS7, cMPS8, cMPS9, cMPS10, cMPS11, cMPS12, cMPS13, cMPS14} can be used to numerically manipulate the cMERA wavefunctional efficiently. Most importantly, these cMPS techniques work equally well for both Gaussian and non-Gaussian wavefunctionals. Therefore they provide a much needed numerical venue for producing strongly correlated (i.e. highly non-Gaussian) cMERA wavefunctionals for interacting QFTs, as demonstrated in Ref. \cite{Ganahl2019}. 

(ii) \textit{Exact ground state of local Hamiltonian:} The magic cMERA $\ket{\Psi^{\Lambda}(s)}$ is the exact ground state of a strictly local QFT Hamiltonian $H^{\Lambda}(s)$, see Eq. \ref{eq:HLams} below. This is unexpected. Indeed, it can be seen that a generic choice of quasi-local entangler produces a wavefunctional that is the ground state of a Hamiltonian which is, at best, quasi-local \cite{Zou2019}. 

Let us elaborate a bit more on this last property, since it will play an important role in our discussion of the $U(1)$ gauge invariant cMERA. 
We introduce the local Hamiltonian
\begin{equation}
    H^{\Lambda} \equiv H^{\CFT} + A^{\Lambda}_{\UV}
\label{eq:HLam}
\end{equation}
for the massless case, where
\begin{equation}
A^{\Lambda}_{\UV} \equiv  \dfrac{1}{2\Lambda^2}\int dx\;(\partial_x\pi(x))^2,
\end{equation}
and the local Hamiltonian
\begin{equation}
    H^{\Lambda}(s) \equiv H_{m(s)}+ A^{\Lambda}_{\UV},
\label{eq:HLams}    
\end{equation}
in the massive case, where $H_{m(s)}$ is the relativistic massive Hamiltonian $H_m$ of Eq. \eqref{eq:Hm} for mass $m(s)$ given by \eqref{eq:mass}. Then Ref. \cite{Zou2019} showed that $\ket{\Psi^{\Lambda}}$ is the exact ground state of $H^{\Lambda}$ and $\ket{\Psi^{\Lambda}(s)}$ is the exact ground state of $H^{\Lambda}(s)$.

The parent Hamiltonian $H^{\Lambda}(s)$ is thus obtained from the relativistic massive $H_{m(s)}$ by adding the non-relativistic UV regulator $A_{\UV}^{\Lambda}$, which breaks Lorentz invariance and primarily affects the UV physics by modifying the dispersion relation for momenta above the cutoff scale $\Lambda$. On the other hand, the mass term in $H^{\Lambda}(s)$ introduces a mass gap in the low energy spectrum. Thus, for $s>0$, that is $m < \Lambda$, we can think of $m$ and $\Lambda$ in $H^{\Lambda}(s)$ as providing IR and UV regulators to the relativistic, massless Hamiltonian $H^{\CFT}$, respectively. 

\subsection{U(1) gauge theory} 

Next we summarize the quantization of a \textit{gauge invariant}, \textit{massless} vector boson field, which is nontrivial due to the presence of constraints (then below we will also discuss the \textit{massive} case, which is no longer gauge invariant). Our goal is to remind the reader of the characterization of the ground state in terms of annihilation operators, which will be useful in the forthcoming analysis. For a more detailed review of the quantization procedure see Appendix \ref{app:constraint}.
\par Consider the free Maxwell Lagrangian for a bosonic vector field $A_\mu$ in $d+1$ spacetime dimensions,
\begin{equation}
\mc L \equiv  -\dfrac{1}{4}F_{\mu\nu}F^{\mu\nu},
\label{eq:lagrangian}
\end{equation}
where $F_{\mu\nu} \equiv  \partial_\mu A_\nu-\partial_\nu A_\mu$ is the field strength tensor. We will quantize the theory in the temporal gauge $A^0=0$. The spatial components $A_i$ of the vector field and their corresponding conjugate momenta $\Pi_i$ are promoted to operators satisfying the canonical commutation relations
\begin{equation}
[A_i(\vx),\Pi_j(\vy)]=i\delta(\vx-\vy)
\label{eq:ccr}
\end{equation}
Due to gauge invariance, physical states are constrained to satisfy Gauss's law
\begin{equation}
\partial_i\Pi^i(\vx)\ket{\text{phys}} = 0,~~~~\forall \vx \in \mathbb{R}^d.
\label{eq:constraint_real}
\end{equation}
The Hamiltonian of the theory is given by
\begin{align}
H_{\gauge}&\equiv  \ha\int{d^dx\;\left[\Pi_i(\vx)\Pi^i(\vx)\right.}\nonumber\\&~~~~~~~~~~~~~-A^i(\vx)\left(\delta_{ij}\Delta -\partial_i\partial_j\right) A^j(\vx)\left.\hspace{-1mm}\right]\label{eq:Hposition}
\\&=\ha\int{d^dk\;\left[\Pi_i(-\vk)\Pi^i(\vk)\right.}\nonumber\\&~~~~~~~~~~~~~+A^i(-\vk)\left(\delta_{ij}k^2  -k_ik_j\right) A^j(\vk)\left.\hspace{-1mm}\right],
\end{align}
where $k=|\vk|$. $H$ can be diagonalized by changing to a basis consistent of the \textit{longitudinal} polarization
\begin{equation}
A_\parallel(\vk)=\dfrac{ik^iA_i(\vk)}{k},\quad \Pi_\parallel(\vk)=\dfrac{ik^i\Pi_i(\vk)}{k} \label{eq:pol1}
\end{equation}
and the $d-1$ orthogonal \textit{transversal} polarizations \footnote{There is a subtlety here involving the fact that the zero mode subspace of the operator algebra, generated by $A_i(\vk = 0),\Pi_i(\vk = 0)$ cannot be separated in longitudinal and transversal sectors. Since this will however not be very relevant to the discussion, we will mostly ignore the zero modes in this paper.}
\begin{equation}
A_{\perp,n}(\vk),~~~\Pi_{\perp,n}(\vk),\qquad n=1,\ldots,d-1. \label{eq:pol2}
\end{equation}
(Summation over any repeated index $n$ for this basis will be implied throughout this paper.) This basis is additionally helpful because the longitudinal polarization is precisely the gauge degree of freedom, while the transversal polarizations are the physical (gauge invariant) degrees of freedom, as can be seen by performing a gauge transformation:
\begin{align}
&A_j(\vx)\to A_j(\vx)+\partial_j\omega(\vx)\nonumber\\
&\implies\begin{cases}
A_\parallel(\vk)&\to A_\parallel(\vk)-k\omega(\vk),\\
A_{\perp,n}(\vk)&\to A_{\perp,n}(\vk).
\end{cases}
\end{align}
Consequently, the gauge constraint \eqref{eq:constraint_real} becomes
\begin{equation}
\Pi_\parallel(\vk)\ket{\text{phys}}=0,~~~\forall \vk \in \mathbb{R}^d.
\label{eq:gauge_constraint_momentum}
\end{equation}
The Hamiltonian $H_{\gauge}$ restricted to the gauge invariant subspace reads
\begin{align}
H_{\gauge}&=\int{d^dk\;\left( \Pi_{\perp,n}(-\vk)\Pi_{\perp,n}(\vk)\right.}\nonumber\\&~~~~~~~~~~~~~~~~~~~~~~~~\left. +k^2A_{\perp,n}(-\vk)A_{\perp,n}(\vk)\right) \\
&=\int{d^dk\;k\; a_{\perp,n}^\dagger(\vk)a_{\perp,n}^{\phantom{\dagger}}(\vk)},
\end{align}
where 
\begin{equation} \label{eq:aU1}
a_{\perp,n}(\vk) \equiv  \sqrt{\dfrac{k}{2}}A_{\perp,n}(\vk)+i\sqrt{\dfrac{1}{2k}}\Pi_{\perp,n}(\vk),
\end{equation}
and we have removed an infinite constant term from the Hamiltonian in the usual way. What remains is nothing but the Hamiltonian for $d-1$ free bosons, whose ground state $\ket{\Psi_{\gauge}}$ is defined via the annihilation operators:
\begin{equation}
a_{\perp,n}(\vk)\ket{\Psi_{\gauge}}=0\qquad\forall\vk,\;n=1,\ldots, d-1.
\label{eq:non_gauge_constraints}
\end{equation}
Notice that both Eq. \eqref{eq:non_gauge_constraints} and the gauge constraint \eqref{eq:gauge_constraint_momentum} are constraints that are expressed in terms of operators that are linear in the field operators $A_i(\vec{x})$ and $\Pi_i(\vec{x})$. These constraints completely determine the ground state $\ket{\Psi_{\gauge}}$, which is therefore a Gaussian state.  

For later reference, we parameterize the annihilation operators $a_{\parallel}(\vk)$ and $a_{\perp}(\vk)$ in terms of two functions $\alpha_\parallel(k), \alpha_\perp(k)$ and write:
\begin{align}
&a_\parallel(\vk)\ket{\Psi_{\gauge}}= 0,~~~~~a_{\perp,n}(\vk)\ket{\Psi_{\gauge}}=0,~~~~\label{eq:general1}\\
&a_{\parallel}(\vk) \equiv  \sqrt{\dfrac{\alpha_\parallel(k)}{2}}A_{\parallel}(\vk)+i\sqrt{\dfrac{1}{2\alpha_\parallel(k)}}\Pi_{\parallel}(\vk),\label{eq:general2}\\
&a_{\perp,n}(\vk) \equiv  \sqrt{\dfrac{\alpha_\perp(k)}{2}}A_{\perp,n}(\vk)+i\sqrt{\dfrac{1}{2\alpha_\perp(k)}}\Pi_{\perp,n}(\vk).\label{eq:general3}
\end{align}

We see that in order to recover  \eqref{eq:gauge_constraint_momentum} and \eqref{eq:non_gauge_constraints} from the more general formulation \eqref{eq:general1}-\eqref{eq:general3}, we just need to make the particular choice of functions $\alpha_\parallel(k), \alpha_\perp(k)$ given by
\begin{equation}
\alpha_\parallel(k)=0,\qquad \alpha_\perp(k)=k\label{eq:massless_alphas},
\end{equation}
where in the particular case of the longitudinal polarization, $\alpha_\parallel(\vk)=0$ implies that the annihilation operator $a_{\parallel}(\vk)$ is proportional to $\Pi_{\parallel}(\vk)$, as required by the gauge constraint.

\subsection{Massive vector boson quantum field theory}

\par For what follows it is also useful to recall the \textit{massive} vector boson theory, obtained by adding a (Proca) mass term to the Lagrangian
\begin{equation}
\mc L_m \equiv  \mc L+\ha m^2A_\mu A^\mu.
\end{equation}
The relevant operator algebra is again generated by \eqref{eq:ccr}, but the mass term breaks gauge invariance so physical states are no longer restricted to satisfy \eqref{eq:gauge_constraint_momentum}. The massive Hamiltonian
\begin{align}
H_m \equiv  H_{\gauge}+\ha\int{d^dx\;\left[ \dfrac{(\partial_i\Pi^i)^2}{m^2}+m^2A_i(x)A^i(x)\right] }
\label{eq:Ham_mass}
\end{align}
can be diagonalized in the polarization basis \eqref{eq:pol1}-\eqref{eq:pol2} as in the massless case, but the lack of gauge invariance implies that the longitudinal component $A_\parallel(\vk)$, $\Pi_\parallel(\vk)$ is now a legitimate propagating degree of freedom, instead of a gauge degree of freedom. The ground state of the theory is again of the form \eqref{eq:general1}-\eqref{eq:general3}, this time with functions $\alpha_\parallel(k), \alpha_\perp(k)$ given by
\begin{equation}
\alpha_\parallel(k)=\dfrac{m^2}{\sqrt{k^2+m^2}},\qquad \alpha_\perp(k)=\sqrt{k^2+m^2}
\label{eq:mass_alphas}
\end{equation}
Note that in the limit $m\to 0$, \eqref{eq:mass_alphas} reduces to \eqref{eq:massless_alphas}. 

\begin{figure}[htp]
	\centering
	\includegraphics[width=6cm]{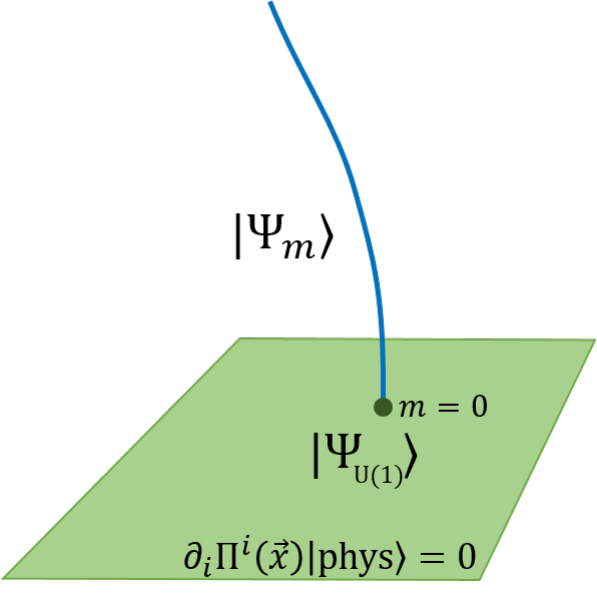}
	\caption{The ground state $\ket{\Psi_m}$ of the massive vector boson Hamiltonian $H_m$ depends on the mass $m$ and is not gauge invariant. However, in the limit $m\rightarrow 0$ we recover massless vector boson Hamiltonian $H_{\gauge}$, whose ground state $\ket{\Psi_{U(1)}}$ is gauge invariant. The green surface represents the gauge invariant subspace, or physical subspace, of the Hilbert space.
	}
	\label{fig:vectorBoson}
\end{figure}

\section{cMERA}
\label{sec3}

We are now ready to present our main result: a cMERA wavefunctional $\ket{\Psi^{\Lambda}_{\gauge}}$ that approximates the ground state $\ket{\Psi_{\gauge}}$  of Hamiltonian $H_{\gauge}$ in Eq. \eqref{eq:Hposition} for the U(1) \textit{gauge invariant}, \textit{massless} free vector boson.

Our construction actually corresponds to a scale invariant cMERA, that is, it is the fixed point of an entangling evolution in scale generated by a constant entangler $K$ starting from an unentangled state $\ket{\Lambda}$ (as introduced earlier in Eq. \eqref{eq:asympt}) and is an extension to gauge fields of the \textit{magic} cMERA of Ref. \cite{Zou2019}. It has the following three key properties:

\vspace{2mm}

(i) \textit{Gauge invariance}: the wavefunctional $\ket{\Psi^{\Lambda}_{\gauge}}$ is explicitly U(1) gauge invariant, that is, it fulfills the constraint \eqref{eq:gauge_constraint_momentum}; 

\vspace{2mm}

\begin{figure}[htp]
	\centering
	\includegraphics[width=\columnwidth]{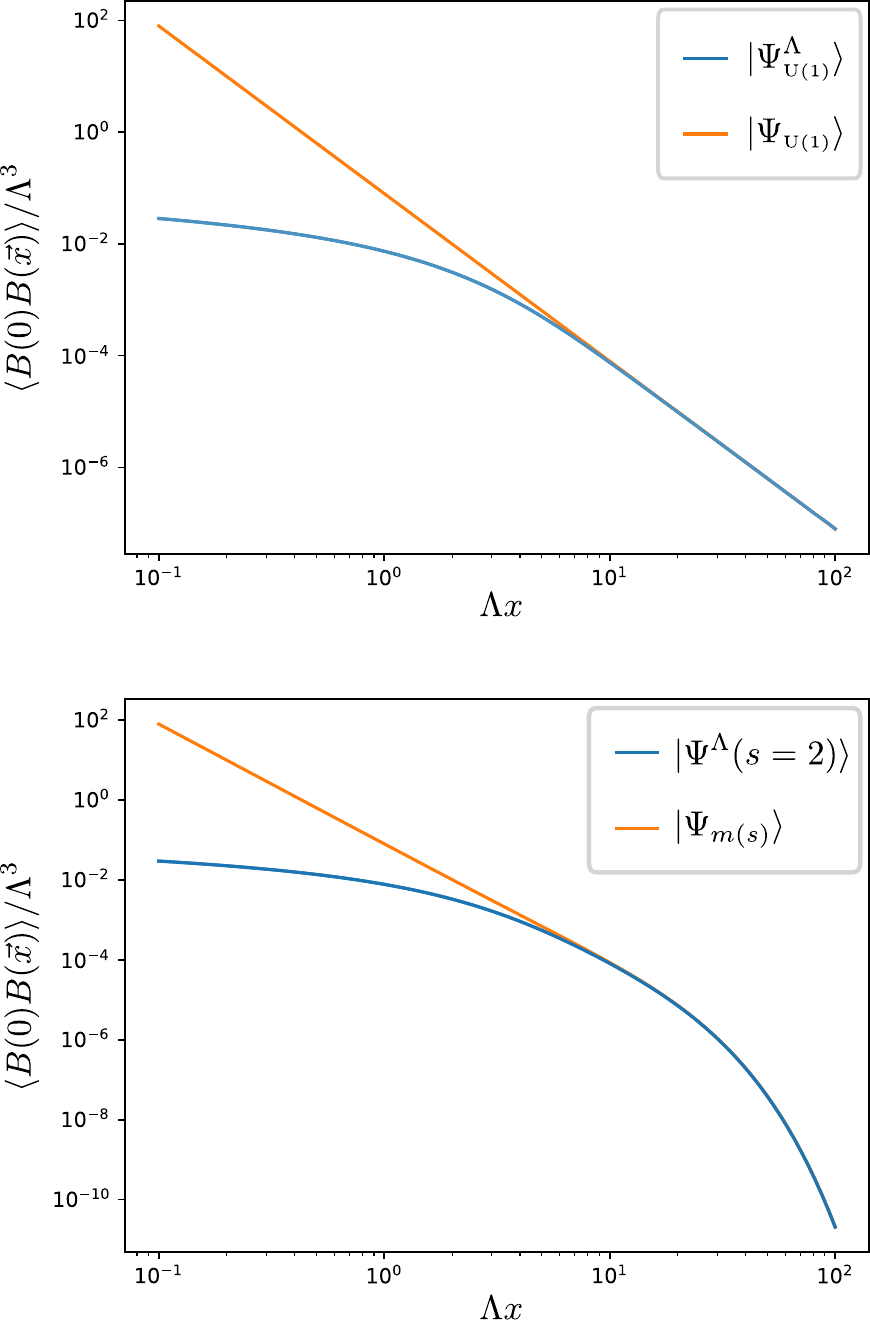}
	\caption{ 
	Correlator $\ev{B(0)B(\vx)}$ as a function of $x \equiv|\vx|$ in 2+1 dimensions, where $B \equiv \partial_1A_2-\partial_2A_1$, for both a target state and the corresponding cMERA approximation, which matches the correlator of its target state for $x\Lambda \gg 1$.
	(top) Gauge invariant target state $\ket{\Psi_{\gauge}}$ and the corresponding cMERA $\ket{\Psi^{\Lambda}_{\gauge}}$. The cMERA correlator has a distributional contribution $\frac{1}{2\Lambda} \left(\Delta+\frac{\Lambda^2}{2}\right) \delta(\vx)$ localized at the origin, not visible in the figure.  
	(bottom) Massive target state $\ket{\Psi_{m(s)}}$ for $m(s) = \Lambda e^{-s}$ and the corresponding cMERA $\ket{\Psi^{\Lambda}(s)}$. The cMERA correlator has a distributional contribution $\frac{1}{2\Lambda} \left(\Delta+\frac{\Lambda^2-m(s)^2}{2}\right) \delta(\vx)$ localized at the origin, not visible in the figure.
	}
	\label{fig:correlators1}
\end{figure}

(ii) \textit{Correct large distance physics (I)}: the wavefunctional $\ket{\Psi^{\Lambda}_{\gauge}}$ accurately approximates the behaviour (e.g. correlators, see Fig. \ref{fig:correlators1}) of the ground state $\ket{\Psi_{\gauge}}$ of $H_{\gauge}$ at distances $x \gg \Lambda^{-1}$, or
\begin{equation}
    \ket{\Psi^{\Lambda}_{\gauge}}~~ \stackrel{\LD}{\sim} ~~\ket{\Psi_{\gauge}}
\end{equation}

\vspace{2mm}

(iii) \textit{Ground state of a local Hamiltonian (I)}: the cMERA $\ket{\Psi^{\Lambda}}$ is the exact ground state of a Hamiltonian $H^{\Lambda}_{\gauge}$, see Eq. \eqref{eq:parentHgauge} below, that is \textit{local} and can be understood as a UV regulated version of $H_{\gauge}$. 

\vspace{2mm}

Moreover, the intermediate cMERA wavefunctional $\ket{\Psi^{\Lambda}(s)}$ for any finite $s \in [0, \infty)$ is related to the massive vector boson described in the previous section in the following ways:

\vspace{2mm}

(iv) \textit{Correct large distance physics (II)}: the wavefunctional $\ket{\Psi^{\Lambda}(s)}$ accurately approximates the behaviour (e.g. correlators, see Fig. \ref{fig:correlators1}) of the ground state $\ket{\Psi_{m(s)}}$ of the relativistic massive vector boson Hamiltonian $H_{m(s)}$, or Hamiltonian $H_m$ in Eq. \eqref{eq:Ham_mass} for mass $m(s) = \Lambda e^{-s}$, that is
\begin{equation}
  \ket{\Psi^{\Lambda}(s)}~~ \stackrel{\LD}{\sim} ~~\ket{\Psi_{m(s)}}.
\end{equation}

(v) \textit{Ground state of a local Hamiltonian (II)}: the cMERA $\ket{\Psi^{\Lambda}(s)}$ is the exact ground state of a Hamiltonian $H^{\Lambda}(s)$, see Eq. \eqref{eq:parentH} below, that is \textit{local} and can be understood as a UV regulated version of $H_{m(s)}$ for mass $m(s) = \Lambda e^{-s}$. 

\vspace{2mm}

\begin{figure}[htp]
	\centering
	\includegraphics[width=6cm]{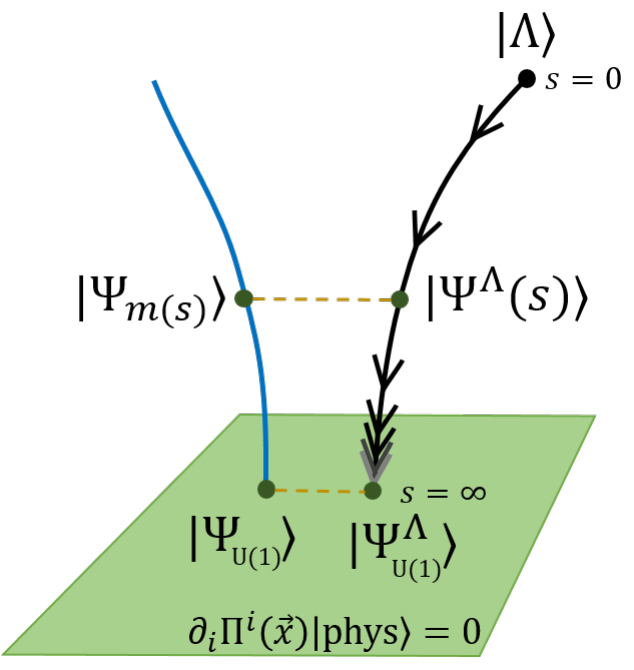}
	\caption{The massive vector boson ground state $\ket{\Psi_{m(s)}}$ for $m(s) = \Lambda e^{-s}$ is approximated by the cMERA state $\ket{\Psi^{\Lambda}(s)}$. None of these wavefunctionals are gauge invariant. The massless vector boson ground state $\ket{\Psi_{\gauge}}$ is approximated by the sacale-invariant cMERA state $\ket{\Psi^{\Lambda}_{\gauge}}$. These two wavefunctionals are gauge invariant.}
	\label{fig:cMERAflow2}
\end{figure}

We emphasize that in our construction, for any finite $s$ (finite mass $m(s)$) the cMERA wavefunctional $\ket{\Psi^{\Lambda}(s)}$ is not gauge invariant, and gauge invariance is only attained in the large $s$ limit. That is, the entangling evolution in scale takes place outside the gauge invariant subspace of the Hilbert space. However, as we will show (see Fig. \ref{fig:correlators2} below), an approximation to $\ket{\Psi^{\Lambda}}$ (e.g. in terms of correlators) can already be obtained from $\ket{\Psi^{\Lambda}(s)}$ at finite $s \gg 1$. Notice that this situation closely mimics the relativistic gauge invariant vector boson we are targetting: at finite mass $m$, the theory is not gauge invariant, and gauge invariance is only attained in the massless limit $m\rightarrow 0$. Fig. \ref{fig:cMERAflow2} summarizes diagramatically the relations between cMERA states and the ground states they target.

\subsection{Unentangled state $\ket{\Lambda}$}

\par We begin by defining annihilation operators
\begin{equation}
\psi_i(\vx)\equiv \sqrt{\dfrac{\Lambda}{2}} A_i(\vx)+i\sqrt{\dfrac{1}{2\Lambda}}\Pi_i(\vx).
\end{equation}
We then consider the unentangled state $\ket{\Lambda}$ given by 
\begin{align}
\psi_i(\vx)\ket\Lambda=0,\quad i=1,\ldots,d.
\label{eq:prod}
\end{align}
This will be the starting point of our entangling evolution. For later convenience, we write \eqref{eq:prod} in the basis of polarizations in momentum space:
\begin{eqnarray}
\psi_\parallel(\vk)\ket\Lambda &=& 0\\
\psi_{\perp, n}(\vk)\ket\Lambda &=& 0,\quad n=1,\ldots,d-1,~~~~~
\label{eq:prodk}
\end{eqnarray}
Here we have defined 
\begin{eqnarray}
\psi_\parallel(\vk)&\equiv& \sqrt{\dfrac{\Lambda}{2}} A_\parallel(\vk)+i\sqrt{\dfrac{1}{2\Lambda}}\Pi_\parallel(\vk),\\
\psi_{\perp,n}(\vk) &\equiv& \sqrt{\dfrac{\Lambda}{2}} A_{\perp,n}(\vk)+i\sqrt{\dfrac{1}{2\Lambda}}\Pi_{\perp,n}(\vk),
\label{eq:psik}
\end{eqnarray}
for $n=1,\cdots,d-1$. The initial state $\ket{\Lambda}$ is clearly also of the Gaussian form \eqref{eq:general1}-\eqref{eq:general3}, with both functions $\alpha_\parallel(k), \alpha_\perp(k)$ set to a constant:
\begin{equation}
\alpha_\parallel(k)=\Lambda,~~~~~\alpha_\perp(k)=\Lambda.
\label{eq:unent_alphas}
\end{equation}

\subsection{Entangling evolution in scale}

Our next step is to define the generator of scale transformations as 
\begin{align}
L&\equiv \ha\int{d^d\vx\;\Pi_i(\vx)\left(-\vx\cdot\vec\nabla-\frac{d}{2}\right) A^i(\vx)+\text{h.c.}}\label{eq:L}\\
&=\int{d^d\vk\;\Pi_i(-\vk)\left(\vk\cdot\vec\nabla_{\vk}+\frac{d}{2}\right) A^i(\vk)+\text{h.c.}},
\end{align}
which assigns non-relativistic scaling dimensions $\Delta_{A_i}=d/2$ and $\Delta_{\Pi_i}=d/2$ to the fields,  
and consider an entangler of the form
\begin{align}
K&=\frac{-i}{2}\int{d^d\vx\;d^d\vy\;g^{ij}(\vx-\vy)\; \psi^{\phantom{\dagger}}_i(\vx)\psi_j^{\phantom{\dagger}}(\vy)+\text{h.c.} }\\
&=\frac{-i}{2}\int{d^d\vk\;g^{ij}(\vk)\; \psi^{\phantom{\dagger}}_i(-\vk)\psi_j^{\phantom{\dagger}}(\vk)+\text{h.c.}}
\end{align}
which is the natural generalization to vector bosons of the scalar boson entangler \eqref{eq:entangler}. We choose a rotation covariant form for the profile $g^{ij}(\vk)=g(k)\delta^{ij}+f(k)k^{i}k^{j}$, and rewrite
\begin{align}
K=\frac{-i}{2}\int{d^d\vk\;\left[ g_\parallel(k)\; \psi^{\phantom{\dagger}}_\parallel(-\vk)\psi^{\phantom{\dagger}}_\parallel(\vk)\ \right.}~~~~~~~~~~~~~~\nonumber\\+\left. g_\perp(k) \;\psi^{\phantom{\dagger}}_{\perp, n}(-\vk)\psi^{\phantom{\dagger}}_{\perp, n}(\vk)\right] +\text{h.c.}
\label{eq:formofK}
\end{align}
where we have defined
\begin{equation}
g_\perp(k)\equiv g(k),\qquad g_\parallel(k)\equiv g(k)+k^2f(k).
\end{equation}
In our case, we choose 
\begin{equation}
g_\parallel(k)=1-\ha\dfrac{\Lambda^2}{\Lambda^2+k^2},\qquad g_\perp(k)=\ha\dfrac{\Lambda^2}{\Lambda^2+k^2}.
\label{eq:ourgs}
\end{equation}
To have a picture of what these profiles look like in position space, notice that by inverse Fourier transforming we obtain
\begin{equation}
\mc F^{-1}\left[\dfrac{1}{\Lambda^2+k^2}\right](x)\propto\dfrac{K_{\frac{d-2}{2}}(\Lambda x)}{(\Lambda x)^{\frac{d-2}{2}}},
\end{equation}
where $x=|\vx|$ and $K_n$ is the $n$-th modified Bessel function of the second kind. This implies that the position space profile of the entangler decays exponentially at large distances, and for $d>1$ it diverges at the origin. In particular, in 1+1 spacetime dimensions, $g_\perp(x)$ is the same profile as that in Eq. \eqref{eq:magicg}.
\par Having made our choices for $\ket{\Lambda}, L$ and $K$, the family of ansatz states $\ket{\Psi^\Lambda(s)}$ is defined via Eq. \eqref{eq:evol}. Since $K$ is a quadratic operator in the fields, and we start from a Gaussian state $\ket\Lambda$, the whole evolution takes place in the manifold of Gaussian states, and each $\ket{\Psi^\Lambda(s)}$ is of the form \eqref{eq:general1}-\eqref{eq:general3}, i.e., it is given by a set of scale-dependent annihilation operators
\begin{eqnarray}
a^\Lambda_{\parallel}(\vk,s)\ket{\Psi^\Lambda(s)}&=&0,~~ \forall \vk\in \R^d,\label{eq:fullcMERA1}\\
a^\Lambda_{\perp,n}(\vk,s)\ket{\Psi^\Lambda(s)}&=&0, ~~\forall \vk\in \R^d,~~ n=1,\cdots, d-1,~~~~~\label{eq:fullcMERA2}
\end{eqnarray}
that are characterized by a pair of scale-dependent functions $\alpha_\parallel(k,s)$, $\alpha_\perp(k,s)$,
\begin{align}
a^\Lambda_{\parallel}(\vk,s)&=\sqrt{\dfrac{\alpha_\parallel(k,s)}{2}}A_{\parallel}(\vk)+i\sqrt{\dfrac{1}{2\alpha_\parallel(k,s)}}\Pi_{\parallel}(\vk),\label{eq:anni1}\\
a^\Lambda_{\perp,n}(\vk,s)&=\sqrt{\dfrac{\alpha_\perp(k,s)}{2}}A_{\perp,n}(\vk)+i\sqrt{\dfrac{1}{2\alpha_\perp(k,s)}}\Pi_{\perp,n}(\vk).\label{eq:anni2}
\end{align}
Using Eq. \eqref{eq:evol} we can solve for $\alpha_\parallel(k,s)$ and $\alpha_\perp(k,s)$ in terms of $g_\parallel(k,s)$ and $g_\perp(k,s)$:
\begin{align}
\alpha_\parallel(k,s)&=\Lambda\exp{\left(-2\int_0^s{du\;g_\parallel(ke^{s-u})}\right)},\\\alpha_\perp(k,s)&=\Lambda\exp{\left(-2\int_0^s{du\;g_\perp(ke^{s-u})}\right)}.
\end{align}
For our particular choice of entangler given by Eq \eqref{eq:ourgs}, we have
\begin{align}
\alpha_\parallel(k,s)&=\dfrac{m(s)^2}{\Lambda}\sqrt{\dfrac{k^2+\Lambda^2}{k^2+m(s)^2}},\label{eq:alfa1}\\
\alpha_\perp(k,s)&=\Lambda\sqrt{\dfrac{k^2+m(s)^2}{k^2+\Lambda^2}},\label{eq:alfa2}
\end{align}
where $m(s)=\Lambda e^{-s}$. Since for the transversal modes of the vector boson we used the same entangler as the one for a scalar boson in Ref. \cite{Zou2019}, $\alpha_\perp(k,s)$ is the same function as $\alpha(k,s)$ in Ref. \cite{Zou2019}.

\subsection{Fixed-point wavefunctional and gauge invariance}

\par In the limit $s\to\infty$, the constraint from Eq. \eqref{eq:fullcMERA1} becomes the gauge invariance condition \eqref{eq:gauge_constraint_momentum}, so that the fixed-point state $\ket{\Psi^\Lambda_{\gauge}}$ belongs to the gauge invariant subspace. It is fully characterized (up to a global phase) by the gauge constraint and the ${s\to\infty}$ limit of the annihilation operators of the transversal modes:
\begin{align}
\Pi_\parallel(\vk)\ket{\Psi^\Lambda}&=0,
\label{eq:cMERAstate1}\\
a^\Lambda_{\perp,n}(\vk, \infty)\ket{\Psi^\Lambda}&=0.\label{eq:cMERAstate2}
\end{align}
The state defined by the conditions \eqref{eq:cMERAstate1}-\eqref{eq:cMERAstate2} is a fixed point of the evolution generated by $L+K$. This can be shown in the same way as it was shown in Ref. \cite{Qi} for the scalar boson.
There is a subtlety regarding the $s\rightarrow \infty$ limit since the theories for $s<\infty$ and $s=\infty$ are fundamentally distinct. We elaborate on this last point in Appendix \ref{app:1d}.

\subsection{Comparison of Gaussian wavefunctionals}

The fact that all states involved in this discussion are Gaussian, of the form \eqref{eq:general1}-\eqref{eq:general3}, facilitates comparison among them, since it can be conducted at the level of $\alpha$ functions.
It was argued in \cite{Qi} that annihilation operators for noninteracting cMERA states interpolate between those of the target state at small momenta $k\ll\Lambda$ and those of the unentangled initial state at large momenta $k\gg\Lambda$. As numerically checked in \cite{Adrian}, this leads to correlation functions with the corresponding interpolating behaviours. Indeed, the two-point functions of $\ket{\Psi^\Lambda(s)}$, which for Gaussian states encode all the other correlators, are intimately related to $\alpha_\parallel(k,s)$ and $\alpha_\perp(k,s)$:
\begin{align}
\ev{A_\parallel(\vk)A_\parallel(\vq)}=\ha\dfrac{\delta(\vk+\vq)}{\alpha_\parallel(\vk,s)},\\
\ev{A_{\perp,n}(\vk)A_{\perp,n}(\vq)}=\ha\dfrac{\delta(\vk+\vq)}{\alpha_\perp(\vk,s)}.
\label{eq:two-point}
\end{align}
In the particular case of our current proposal, it follows from Eqs. \eqref{eq:alfa1}-\eqref{eq:alfa2} that
\begin{align}
\label{eq:aparalimits}\alpha_\parallel(k,s)\sim\begin{cases}
\dfrac{m(s)^2}{\sqrt{k^2+m(s)^2}} & k\ll\Lambda,\\\\
\dfrac{m(s)^2}{\Lambda} & k\gg\Lambda,
\end{cases}\\\nonumber\\\label{eq:aperplimits}
\alpha_\perp(k,s)\sim\begin{cases}
\sqrt{k^2+m(s)^2} & k\ll\Lambda,\\
\Lambda & k\gg\Lambda.
\end{cases}
\end{align}
We see that for $k\ll\Lambda$ these functions reproduce the target state's behaviour (see Eq. \eqref{eq:mass_alphas}), while for $k\gg\Lambda$ they become constant, which is the behaviour seen for the unentangled initial state $\ket{\Lambda}$ (see Eq. \eqref{eq:unent_alphas}. The longitudinal case is special since the constant is rescaled along the evolution from $\Lambda$ at $s=0$ to 0 at $s=\infty$). 

\begin{figure}[htp]
	\centering
	\includegraphics[width=\columnwidth]{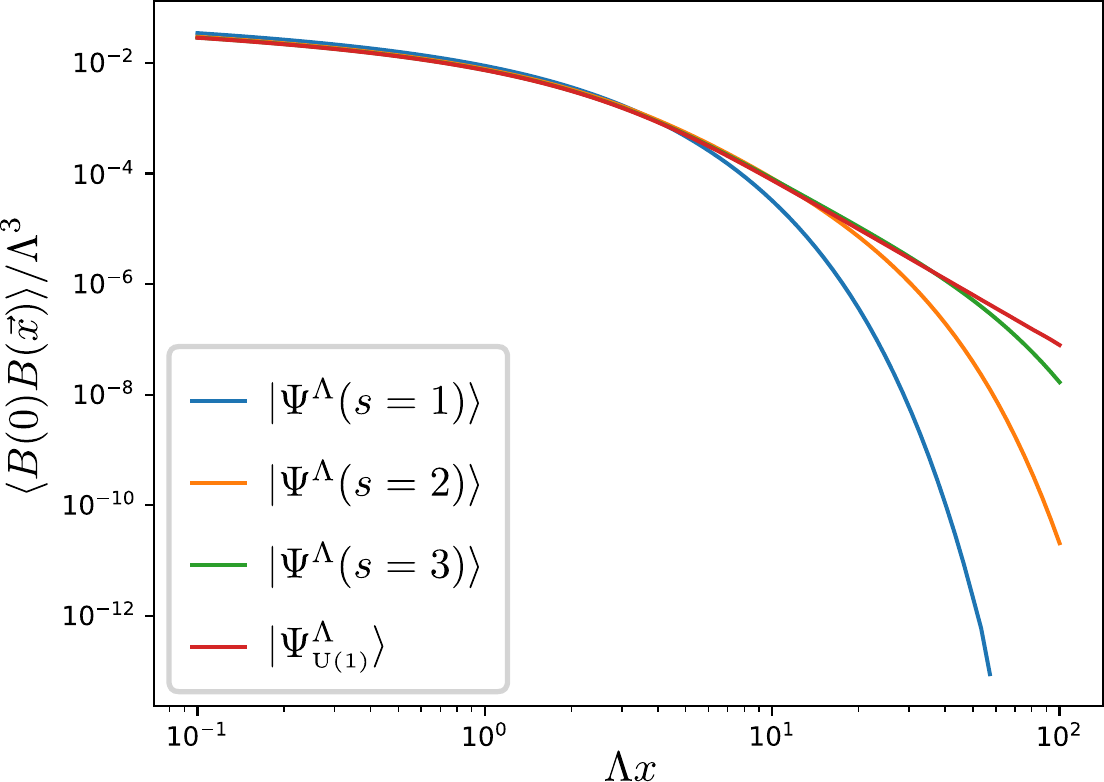}
	\caption{Two-point correlator $\ev{B(0)B(\vx)}$ for state $\ket{\Psi^{\Lambda}(s)}$ for $s=1,2,3$ and for the fixed point wavefunctional $\ket{\Psi^{\Lambda}_{\gauge}}$ in 2+1 dimensions (see also Fig. \ref{fig:correlators1}).
	}
	\label{fig:correlators2}
\end{figure}

Fig. \ref{fig:correlators2} shows the correlator $\ev{B(0)B(\vx)}$ for $\ket{\Psi^{\Lambda}(s)}$ as a function of $s$. We see that, aside from a delta at the origin (see also Appendix B), the correlators for large $s$ converge to those of the fixed point wavefunctional $\ket{\Psi^{\Lambda}_{\gauge}}$. Thus, we can learn about the proporties of the gauge invariant $\ket{\Psi^{\Lambda}_{\gauge}}$ by studying the non-gauge invariant $\ket{\Psi^{\Lambda}(s)}$ at finite but large $s$. 

\subsection{Local Hamiltonians with relativistic IR physics}

\par In order to prove statements (iii) and (v) from the beginning of this section, consider the following family of Hamiltonians:
\begin{equation}
H^{\Lambda}(s) \equiv H_{m(s)}+ B^{\Lambda}_{\UV}(s)  
\label{eq:parentH}
\end{equation}
where $H_{m(s)}$ is the massive Hamiltonian from \eqref{eq:Ham_mass} with ${m(s)=\Lambda e^{-s}}$ and 
\begin{align}
B^{\Lambda}_{\UV}(s) \equiv  \dfrac{1}{\Lambda^2}&\int{d^dx\;\Pi^i(\delta_{ij}\Delta-\partial_i\partial_j)\Pi^j}\nonumber\\+\dfrac{m(s)^2}{\Lambda^2}&\int{d^dx\;(\partial_iA^i)^2}.
\label{eq:cutoffHamiltonian}
\end{align}
For every $s\in[0,\infty]$, $H^{\Lambda}(s)$ is quadratic and can hence be easily diagonalized. We then find that $\ket{\Psi^\Lambda(s)}$ is the ground state of $H^{\Lambda}(s)$. The term $B^{\Lambda}_{\UV}(s)$ in \eqref{eq:parentH} can be seen as a UV regulator for $H_{m(s)}$. Notice that the first line in \eqref{eq:cutoffHamiltonian} involves the transversal degrees of freedom, while the second line involves the longitudinal one:
\begin{align}
B^{\Lambda}_{\UV}(s)=\dfrac{1}{\Lambda^2}&\int{d^dk\;k^2\,\Pi_{\perp,n}(-\vk)\Pi_{\perp,n}(\vk)}\nonumber\\+\dfrac{m(s)^2}{\Lambda^2}&\int{d^dk\;k^2A_\parallel(-\vk)A_\parallel(\vk)}.
\label{eq:cutoff_momentum}
\end{align}
The UV regulator term for the transversal modes is once again equivalent to the one found in \cite{Zou2019}. In the limit $s\to\infty$, the longitudinal degree of freedom is restricted by the gauge constraint, and $\ket{\Psi^\Lambda_{\gauge}}$ can be given a rather compact parent Hamiltonian:
\begin{equation}
H_{\gauge}^{\Lambda} \equiv H_{\gauge} + \dfrac{1}{\Lambda^2}\int{d^dx\;\Pi_i(x)\Delta\Pi^i(x)}.
\label{eq:parentHgauge}
\end{equation}
with $H_{\gauge}$ the Maxwell Hamiltonian from \eqref{eq:Hposition}.

\section{Discussion}

In this paper we have explained, through the concrete example of U(1) gauge theory (or electromagnetism without matter fields) in $d+1$ spacetime dimensions, how to extend the cMERA formalism to gauge invariant quantum fields. In particular, we have seen that the gauge constraint can be made compatible with the UV structure of the cMERA wavefunctional.  As with previous cMERA constructions, the ansatz state can be understood as a the result of modifying the short-distance structure of the target state. Additionally, the resulting cMERA wavefuctional $\ket{\Psi^{\Lambda}(s)}$ has been seen to be the exact ground state of a local Hamiltonian, obtained from the original relativistic Hamiltonian by adding a non-relativistic term that modifies its UV behaviour. 

As in the case of the free scalar boson of Ref. \cite{Zou2019}, for finite $s$ the entangling evolution in scale produces  
a cMERA ansatz for the massive theory with mass $m(s)=\Lambda e^{-s}$. However, for the vector boson analysed here this came with an interesting twist: since a mass term breaks gauge invariance, the intermediate cMERA state $\ket{\Psi^{\Lambda}(s)}$ is not in the physical subspace of the gauge theory, and the whole cMERA evolution happens outside of it, with only the asymptotic fixed point cMERA state $\ket{\Psi^{\Lambda}_{\gauge}}$ being gauge invariant. 

Our ultimate goal is to use cMERA to simulate interacting gauge theories in $d+1$ spacetime dimensions. While for $d=1$ non-Gaussian cMERAs for interacting QFTs can be numerically manipulated using cMPS techniques \cite{Ganahl2019}, no analogous strategy has yet been developed for $d>1$.

\section*{Acknowledgements}
We thank Yijian Zou for enlightening discussions, and Stefan K\"uhn for his guidance through the lattice gauge theory literature. We also thank Martin Ganahl and Qi Hu for providing feedback on the manuscript. A. Franco-Rubio thanks the Quantum Information and String Theory program at the Yukawa Institute for Theoretical Physics (Kyoto), the Quantum Information workshop at the Centro de Ciencias de Benasque Pedro Pascual, the Quantum Information Theory research term at the Instituto de F\'isica Te\'orica and the Instituto de Ciencias Matem\'aticas (Madrid), and the Max Planck Institute for Quantum Optics (Munich), where parts of this work were completed. G. Vidal is a CIFAR fellow in the Quantum Information Science Program. X is formerly known as Google[x] and is part of the Alphabet family of companies, which includes Google, Verily, Waymo, and others (www.x.company). Research at Perimeter Institute is supported by the Government of Canada through the Department of Innovation, Science and Economic Development Canada and by the Province of Ontario through the Ministry of Research, Innovation and Science.
\appendix
\section{Review of the quantization of theories with constraints}
\label{app:constraint}
The field theories we deal with in this paper are special in the sense that they involve constraints due to the Lagrangian not being regular. There are several nuances that should be taken into account when quantizing a constrained system, and in general there might not be a unique way of doing so. For instance, the Maxwell Lagrangian presents gauge invariance, and can thus be quantized by fixing the gauge in a variety of ways. Here we choose a quantization scheme that makes the massive and massless theories ``compatible'', in a sense that we will specify below. The motivation for this choice comes from the magic cMERA for a scalar field \cite{Zou2019}, where the evolution in scale that asymptotically generates the cMERA can be interpreted as the removal of an IR cutoff given by a mass (see main text). In this Appendix we review the canonical quantization procedure we make use of in the main text. For a more thorough explanation we refer the reader to a specialized textbook such as \cite{gaugeBook}. From now on, $\{,\}$ refers to the canonical Poisson bracket, and we usually omit the space dependence of the fields.
\par These are the general steps we will follow for the vector boson theories:
\begin{enumerate}
	\item Given an irregular Lagrangian, find all the constraints to be imposed in the Hamiltonian formalism. This includes primary constraints (dependency relations between coordinates and momenta) and secondary constraints (constraints derived from demanding that other constraints are preserved by the time evolution generated by the Hamiltonian).
	\item Once all constraints have been found, classify them as first-class (if they Poisson-commute with all other constraints) or second-class (if they don't).
	\item Impose second-class constraints as operator equations in the algebra of operators (technically speaking, what we do is replacing Poisson brackets by Dirac brackets when defining the algebra of operators, but for all practical purposes we can think of this step as firstly stated).
	\item For first-class constraints we use two different quantization strategies: either we impose them as gauge invariance constraints on the Hilbert space (claiming only gauge invariant states are physical), or we add an additional constraint to make them second class, then apply the previous step (gauge fixing).
\end{enumerate}
\par We begin with the massless vector boson (Maxwell) Lagrangian density 
\begin{align}
\mathcal L &=-\frac{1}{4}F_{\mu\nu}F^{\mu\nu}\nonumber\\&= -\frac{1}{4}(\partial_\mu A_\nu-\partial_\nu A_\mu)(\partial^\mu A^\nu-\partial^\nu A^\mu),
\end{align}
and compute the conjugate momenta to the field components in the usual way
\begin{align}
\Pi^0&\equiv \dfrac{\partial\mc L}{\partial(\partial_0A_0)}=0\\
\Pi^i&\equiv \dfrac{\partial\mc L}{\partial(\partial_0A_i)}=-F^{0i}=\partial_0 A_i-\partial_iA_0
\end{align}
We find a single primary constraint: $\Pi^0=0$. This constraint is now included in the Hamiltonian by means of a Lagrange multiplier $u$:
\begin{align}
\mc H = &= \partial_0 A_\mu\Pi^\mu-\mc L + u \Pi^0\\
&=\ha\Pi_i\Pi^i+\dfrac{1}{4}F_{ij}F^{ij}-A_0\partial_i\Pi^i+u\Pi^0
\end{align}
We look for secondary constraints by imposing the preservation of the primary constraint under time evolution:
\begin{equation}
\partial_0\Pi^0=\{\Pi^0,\mc H\}=0\implies\partial_i\Pi^i =0.
\end{equation}
Thus we obtain a secondary constraint, that we identify as Gauss's law. We look for additional secondary constraints and find that
\begin{equation}
\partial_0(\partial_i\Pi^i)=\{\partial_i\Pi^i,\mc H\}=0
\end{equation}
holds without any additional assumptions. Thus we have found all constraints.
\par The two constraints we obtained are first-class since their Poisson bracket vanishes:
\begin{equation}
\{\Pi^0(\vx),\partial_i\Pi^i(\vy)\}=0
\end{equation}
First-class constraints are a consequence of gauge invariance, which it is well-known that the Maxwell field exhibits. 
\par We now make the choice to quantize in the temporal gauge. We add the (partially) gauge-fixing constraint $A_0=0$ (temporal gauge), and check its consistency:
\begin{equation}
\dot A_0 =\{A_0, \mc H\} = \dfrac{\partial \mc H}{\partial\Pi^0}=u.
\end{equation}
Hence we can impose the preservation of the gauge fixing just by a condition on the Lagrange multiplier. The new set of constraints includes a first-class constraint
\begin{equation}
\partial_i\Pi^i=0
\end{equation}
and a pair of second-class constraints
\begin{equation}
	A_0=\Pi^0=0.
\end{equation}
Upon quantization, the second pair of constraints can be imposed (\`{a} la Dirac) as operator equations: the operator representation of $A_0$ and $\Pi^0$ vanishes identically. The remaining first-class constraint generates (``residual'') gauge transformations:
\begin{equation}
A_i\longmapsto A_i+\partial_i\epsilon(x)
\end{equation}
This transformations do not affect the physical degrees of freedom, and any two states that differ by one of them should be identified. Consequently, we define the physical Hilbert space as the subspace of the total Hilbert space whose elements are invariant under $\partial_i\Pi^i$:
\begin{equation}
\partial_i\hat\Pi^i\ket{\text{phys}}=0
\end{equation}
And the relevant operator algebra is just the one generated by the \textit{spatial} components of fields and momenta $A_i,\Pi^i$, with the usual canonical commutation relations:
\begin{equation}
[A_i(\vx),\Pi^j(\vy)]=i\delta_{ij}\delta(\vx-\vy)
\end{equation}
In this setting we can now write down and diagonalize the Hamiltonian operator, as is done in the main text. 
\par We now turn to the massive vector boson, given by the Proca Lagrangian
\begin{equation}
\mc L =-\dfrac{1}{4}F_{\mu\nu}F^{\mu\nu}-\dfrac{1}{2}m^2A_\mu A^\mu.
\end{equation}
The nonvanishing mass term spoils gauge invariance, but this does not mean that the Lagrangian is regular, since we obtain the same primary constraint
\begin{equation}
\Pi^0\equiv \dfrac{\partial\mc L}{\partial(\partial_0A_0)}=0,
\end{equation}
which we include it with a Lagrange multiplier $u$:
\begin{align}
\mc H &= \dot A_\mu\Pi^\mu-\mc L + u \Pi^0\\
&=\ha\Pi_i\Pi^i+\dfrac{1}{4}F_{ij}F^{ij}+\ha m^2A_\mu A^\mu-A_0\partial_i\Pi^i+u\Pi^0.
\end{align}
Looking for secondary constraints, we find
\begin{equation}
\partial_0 \Pi^0 = 0\implies m^2A_0+\partial_i\Pi^i=0,
\end{equation}
and
\begin{equation}
\partial_0(m^2A_0+\partial_i\Pi^i) = 0\implies u=\partial_iA_i,
\end{equation}
thus the theory presents two constraints. These are second-class, since their Poisson bracket does not vanish:
\begin{equation}
\{\Pi^0(\vx),m^2A_0(\vy)+\partial_i\Pi^i(\vy)\}=-m^2\delta(\vx-\vy)
\end{equation}
We proceed to quantize the theory ``\`{a} la Dirac'': we impose the constraints as operator equations, which effectively removes $A_0, \Pi^0$ as independent operators, and impose canonical commutation relations on the rest of operators:
\begin{align}
\Pi^0=0,\;& A_0=-\dfrac{\partial_i\Pi^i}{m^2},\\
[ A_i(\vx),  A_j(\vy)]&=[ \Pi^i(\vx),  \Pi^j(\vy)]=0,\\
[ A_i(\vx), \Pi^j(\vy)]&=\delta_{ij}\delta(\vx-\vy).
\end{align}
As expected, the Hilbert space presents no gauge constraints, unlike the massless case. From here on the Hamiltonian can be reexpressed as in Eq. \eqref{eq:Ham_mass}, and diagonalized by writing it in terms of the corresponding creation-annihilation operators.
 
\par It can be seen now why we have chosen this particular schemes for quantization of the massive and massless vector fields. The relevant operator algebras are identical: we have in both cases $A_i,\Pi_i$, the spatial components of the field and their momenta, and they satisfy canonical commutation relations. (This would not be as trivial if we had chosen, for example, the Coulomb gauge quantization for the gauge theory, where the commutation relations of the operators are modified from the canonical case.) The Hilbert spaces where these observables are represented are also taken to be the same, with the caveat that for the massless case only a subspace of the total Hilbert space is physical, since the number of physical degrees of freedom is reduced by the gauge invariance. This allows for a cMERA evolution to be defined consistently as in the main text, in a way that $\Psi^\Lambda(s)$ is a massive vector boson state for finite $s$ and a massless vector boson state for $s=\infty$.

\section{On the continuity of the $s\to\infty$ limit}
\label{app:1d}
In this Appendix we point out a subtlety with the ${s\to\infty}$ limit of the longitudinal degrees of freedom in the cMERA from the main text. A reader familiar with the $m\to 0$ limit of the massive vector boson theory to the massless vector boson theory will find it analogous to what we present here. To be concrete, we study the particular case of 1+1 dimensions, where there are no transversal degrees of freedom. We thus denote $A(\vk)\equiv  A_1(\vk)=A_\parallel(\vk)$ and $\Pi(\vk)\equiv \Pi_1(\vk)=\Pi_\parallel(\vk)$. The massless case theory is  pure gauge, having no physical degrees of freedom except for the zero mode $A(\vk=0),\Pi(\vk=0)$  (which gives the quantization of the constant value of the electric field, the only physical degree of freedom of the classical theory). The massless Hamiltonian is given by
\begin{equation}
H=\ha\int{dx\;\Pi(x)^2}=\ha\int{dk\;\Pi(-k)\Pi(k)},
\end{equation}
and its ground state is characterized by
\begin{equation}
\Pi(k)\ket{\Psi}=0
\label{eq:exactgroundstate}
\end{equation}
which includes the gauge invariance constraint ($k\neq 0$) and the energy minimization for the single degree of freedom ($k=0$). It is hard to make statements about the entanglement properties of this state, since, even though Fourier transformation of \eqref{eq:exactgroundstate} yields local annihilation operators
\begin{equation}
\Pi(x)\ket{\Psi}=0,
\label{eq:exactgroundstate_position}
\end{equation}
as befits an unentangled state, there are no local physical degrees of freedom to speak about their correlations or lack thereof.
\par
\begin{figure}[htp]
	\centering
	\includegraphics[width=\linewidth]{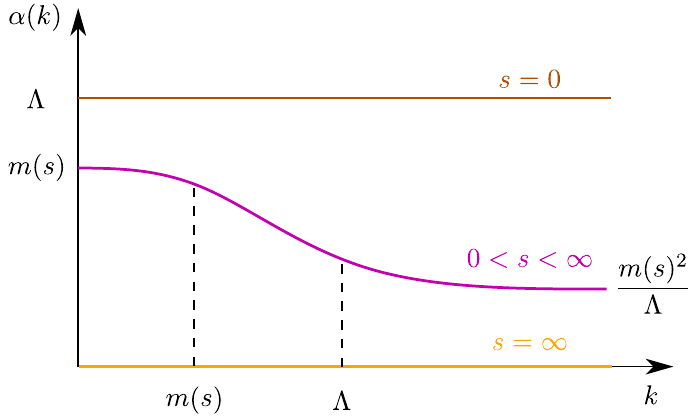}
	\caption{Evolution of $\alpha(k,s)$ (qualitative plot).}
	\label{fig:alpha1p1}
\end{figure}

 Applying the formalism from the main text, we start from the unentangled state
\begin{equation}
\psi(x)\ket{\Lambda}=0
\end{equation}
and evolve with an entangler
\begin{equation}
K = \dfrac{-i}{2}\int{dk\;g(k)\psi(-k)\psi(k)}+\text{h.c.}
\end{equation}
with
\begin{equation}
g(k)=1-\ha\dfrac{\Lambda^2}{\Lambda^2+k^2}.
\end{equation}
In position space, this entangler looks like this
\begin{equation}
K = \int{dx\;A(x)\Pi(x)}-\int{dx\,dy\;e^{-\Lambda|x-y|}A(x)\Pi(y)}
\end{equation}
which has an onsite part, and a bilocal part, which acts at a particular lengthscale $\Lambda^{-1}$.
The cMERA states are then characterized by
\begin{equation}
\left(\sqrt{\dfrac{\alpha(k,s)}{2}}A(k)+i\sqrt{\dfrac{1}{2\alpha(k,s)}}\Pi(k) \right) \ket{\psi^\Lambda(s)}=0
\end{equation}
with
\begin{align}
\alpha(k,s)=\dfrac{m(s)^2}{\Lambda}\sqrt{\dfrac{k^2+\Lambda^2}{k^2+m(s)^2}}.
\end{align}
Figure \ref{fig:alpha1p1} shows a qualitative plot of $\alpha(k,s)$. States at finite values of $s$ are ground states of the following regularized version of the massive vector boson Hamiltonian:
\begin{equation}
H(s) = \int{dx\;\left[ \dfrac{\Pi(x)^2}{2}+\dfrac{m^2}{\Lambda^2}(\partial A(x))^2\right] }
\end{equation}
where $m(s)=\Lambda e^{-s}$. This states are entangled, due to the action of $K$. At any finite time in the evolution, the state has all the entanglement that has been introduced from the UV cutoff scale $\Lambda^{-1}$ to the IR cutoff scale $m(s)^{-1}$. This entanglement does not disappear smoothly in the $s\to\infty$ limit because the convergence to the fixed point is not smooth. Take a look at the two-point function for $\Pi(x)$:
\begin{equation}
\ev{\Pi(x)\Pi(y)}\propto\mathcal{F}^{-1}[\alpha(k)](x-y)
\end{equation}
Because of the behaviour of $\alpha(k)$ we can decompose this correlator into two parts: an onsite delta and an integrable function of the distance:
\begin{equation}
\ev{\Pi(x)\Pi(y)}\sim \frac{m(s)^2}{\Lambda}\delta(x-y)+f_s(x-y)
\end{equation}
As $s\to 0$, we have $\|f_s\|_2\to0$ so both terms in the correlator go to zero. However the value at $x=y$, corresponding to the squared norm of $\Pi(x)\ket{\Psi^\Lambda(s)}$ preserves the delta divergence for the whole evolution, while it should be zero in the gauge invariant subspace.

\section{UV regularization of correlation functions}
\label{app:corr}
In the main text we have provided a cMERA for the ground state of a vector boson theory. The cMERA states can be seen to approximate the long distance properties of their target states, while keeping their short distance properties closer to those of the original unentangled state. In this Appendix we study the UV structure of the proposed gauge invariant cMERA in more detail.

In \cite{Adrian}, the existence of the short distance limit of two-point functions of cMERA states was used as a witness for UV regularization. These correlation functions usually take the following form in cMERA states:
\begin{equation}
\ev{\mc O(\vx)\mc O (\vy)}=C\delta(\vx-\vy)+f(|\vx-\vy|)
\end{equation}
with $C$ a constant and $f$ some function such that
\begin{equation}
\lim_{\vx\to\vy}{\ev{\mc O(\vx)\mc O (\vy)}}<\infty,
\end{equation}
that is, the short-distance limit of the correlator of two fields is finite, barring the on-site delta divergence. In our particular example, and focusing on the $A$ fields, the two-point functions are given in terms of the $\alpha$ functions by \eqref{eq:two-point}. Removing the constant factor responsible for the $\delta$ functions, the corresponding limits are given by
\begin{align}
&\lim_{\vx\to\vy}{\ev{A_\parallel(\vx)A_\parallel(\vy)}}=\int{d^dk\left( \dfrac{1}{\alpha_\parallel(k,s)}-\dfrac{\Lambda}{m(s)^2}\right) }\\
&\lim_{\vx\to\vy}{\ev{A_{\perp, n}(\vx)A_{\perp,n}(\vy)}}=\int{d^dk\left( \dfrac{1}{\alpha_\perp(k,s)}-\dfrac{1}{\Lambda}\right) }
\end{align}
where the expectation values are taken with respect to $\ket{\Psi(k,s)}$. For the $\alpha$ functions from \eqref{eq:alfa1}-\eqref{eq:alfa2} we have
\begin{align}
\dfrac{1}{\alpha_\parallel(k,s)}&\sim \dfrac{\Lambda}{m(s)^2}\left( 1- \dfrac{\Lambda^2 - m(s)^2}{2k^2}+\ldots\right) ,\\
\dfrac{1}{\alpha_\perp(k,s)}&\sim \dfrac{1}{\Lambda}\left(1 + \dfrac{\Lambda^2- m(s)^2}{2 k^2}+\ldots\right),
\end{align}
and the UV divergence is removed in $d=1$, but it remains in higher dimensions. Notice that since 
\begin{equation}
\alpha_\parallel(k,s)=\dfrac{m(s)^2}{\alpha_\perp(k,s)}, 
\end{equation}
and both functions asymptote to constants, their asymptotic behaviour is very much related, as can be seen in the expansions above.
\par If needed, we can build cMERAs where these states are more strongly UV regulated, at the cost of adding extra derivatives to the entangler and the parent Hamiltonians. Consider an entangler of the form \eqref{eq:formofK} whose momentum space profile is given by
\begin{align}
g_\perp(k)&=\frac{1+n \kappa^{2 n-2}}{2 \left(1+\kappa^{2 n-2}\right) \left(1+\kappa^2+\kappa^{2 n}\right)},\quad\kappa\equiv \dfrac{k}{\Lambda},\\
g_\parallel(k)&=1-g_\perp(k)
\end{align}
for $n>1$. $g_\perp(k)$ is a rational function of $k^2$ that goes to $\ha$ at $k=0$ and decays as $k^{-2n}$ at long distances. Its Fourier transform, namely the real space profile of the entangler, is therefore integrable for dimensions $d<2n$. The resulting $\alpha$ functions corresponding to this entangler, if applied on the same initial state, are 
\begin{align}
\alpha_\perp(k,s)&=\Lambda  \sqrt{\frac{k^{2 n}+k^2\Lambda ^{2 n-2}}{k^{2 n}+k^2 \Lambda ^{2 n-2}+\Lambda ^{2 n}}}\,\times\nonumber\\
&~~~~~~~~~~~~\sqrt{\frac{k^{2 n}+k^2 m(s) ^{2 n-2}+m(s) ^{2 n}}{k^{2 n}+k^2m(s) ^{2 n-2}}},\\
\alpha_\parallel(k,s)&= \dfrac{m^2(s)}{\alpha_\perp(k,s)},
\end{align}
with asymptotic fixed-points:
\begin{align}
\alpha_\perp(k)&=\Lambda  \sqrt{\frac{k^{2 n}+k^2\Lambda ^{2 n-2}}{k^{2 n}+k^2 \Lambda ^{2 n-2}+\Lambda ^{2 n}}},\\
\alpha_\parallel(k)&= 0.
\end{align}
These states are more strongly UV regularized, as can be checked by expanding
\begin{align}
\dfrac{1}{\alpha_\parallel(k,s)}&\sim \dfrac{\Lambda}{m(s)^2} + \dfrac{\Lambda \left(m(s)^{2n} - \Lambda^{2n}\right)}{2 m(s)^2k^{2n}}+\ldots,\\
\dfrac{1}{\alpha_\perp(k,s)}&\sim \dfrac{1}{\Lambda} + \dfrac{\Lambda^{2n}- m(s)^{2n}}{2 \Lambda k^{2n}}+\ldots,
\end{align}
Indeed, the short distance limit of the corresponding two-point functions will be finite for $d<2n$, while their long distance behaviour can still be seen to approximate that of the target state correlation functions.


\begin{thebibliography}{99}


	\bibitem{weinberg}
	S.~Weinberg,
	\textit{The quantum theory of fields. Vol. 2: Modern applications}, 1996
	
	\bibitem{schwartz}
	M.~D.~Schwartz,
	\textit{Quantum Field Theory and the Standard Model}, 2013
	
	\bibitem{misner} 
	C.~W.~Misner, K.~S.~Thorne, J.~A.~Wheeler,
	\textit{Gravitation}, 1973
	
	\bibitem{wen}
	X.-G.~Wen,
	\textit{Quantum Field Theory of Many-Body Systems}, 2004

	\bibitem{gaugeBook} 
	M.~Henneaux and C.~Teitelboim,
	\textit{Quantization of gauge systems},
	(Princeton University Press, 1994)
	
	\bibitem{Wilson} 
	K.~G.~Wilson,
	\textit{Confinement of Quarks},
	Phys.\ Rev.\ D {\bf 10}, 2445 (1974).
	doi:10.1103/PhysRevD.10.2445
	
	\bibitem{KogutSusskind} 
	J.~B.~Kogut and L.~Susskind,
	\textit{Hamiltonian Formulation of Wilson's Lattice Gauge Theories},
	Phys.\ Rev.\ D {\bf 11}, 395 (1975).
	doi:10.1103/PhysRevD.11.395
	
	\bibitem{QCD1} 
	S.~Aoki {\it et al.} [CP-PACS Collaboration],
	\textit{Quenched light hadron spectrum},
	Phys.\ Rev.\ Lett.\  {\bf 84}, 238 (2000)
	doi:10.1103/PhysRevLett.84.238
	[hep-lat/9904012].
	
	\bibitem{QCD2} 
	S.~Durr {\it et al.},
	\textit{Ab-Initio Determination of Light Hadron Masses},
	Science {\bf 322}, 1224 (2008)
	doi:10.1126/science.1163233
	[arXiv:0906.3599 [hep-lat]].
	
	\bibitem{sign1}
	C.~Gattringer and K.~Langfeld,
	\textit{Approaches to the sign problem in lattice field theory},
	International Journal of Modern Physics A, Vol. 31, No. 22, 1643007 (2016)
	doi:10.1142/S0217751X16430077
	[arXiv:1603.09517]
	
	\bibitem{sign2}
	D.~Sexty,
	\textit{New algorithms for finite density QCD}, in proceedings of the 32nd International Symposium on Lattice Field Theory 
	PoS(LATTICE2014)016
	doi:10.22323/1.214.0016
	
	\bibitem{MERA1} 
	G. Vidal, 
	\textit{Entanglement renormalization}, 
	Phys. Rev. Let. 99, 220405 (2007),  
	arXiv:cond-mat/0512165. 
	
	\bibitem{MERA2} 
	G. Vidal,
	\textit{A class of quantum many-body states that can be efficiently simulated}, 
	Phys. Rev. Lett. 101, 110501 (2008), 
	arXiv: quant-ph/0610099.
	
	\bibitem{Swingle} 
	B.~Swingle,
	\textit{Entanglement Renormalization and Holography},
	Phys.\ Rev.\ D {\bf 86}, 065007 (2012)
	doi:10.1103/PhysRevD.86.065007
	[arXiv:0905.1317 [cond-mat.str-el]].
	
	\bibitem{Czech} 
	B.~Czech, L.~Lamprou, S.~McCandlish and J.~Sully,
	\textit{Tensor Networks from Kinematic Space},
	JHEP {\bf 1607}, 100 (2016)
	doi:10.1007/JHEP07(2016)100
	[arXiv:1512.01548 [hep-th]].
	
	%Hamiltonian 1+1d %%%%%%%%%%%%%%%%%%%%%%%%%%%%%%%%%%
	
	\bibitem{LGTTN1Ham1} 
	T.M.R.~Byrnes, P.~Sriganesh, R.~J.~Bursill and C.~J.~Hamer,
	\textit{Density matrix renormalization group approach to the massive Schwinger model},
	Phys.\ Rev.\ D {\bf 66}, 013002 (2002)
	doi:10.1103/PhysRevD.66.013002
	[hep-lat/0202014].
	
	\bibitem{LGTTN1Ham2} 
	T.~Sugihara,
	\textit{Matrix product representation of gauge invariant states in a Z(2) lattice gauge theory},
	JHEP {\bf 0507}, 022 (2005)
	doi:10.1088/1126-6708/2005/07/022
	[hep-lat/0506009].
	
	\bibitem{LGTTN1Ham3} 
	M.~C.~Ba\~nuls, K.~Cichy, K.~Jansen and J.~I.~Cirac,
	\textit{The mass spectrum of the Schwinger model with Matrix Product States},
	JHEP {\bf 1311}, 158 (2013)
	doi:10.1007/JHEP11(2013)158
	[arXiv:1305.3765 [hep-lat]].
	
	\bibitem{LGTTN1Ham4} 
	B.~Buyens, J.~Haegeman, K.~Van Acoleyen, H.~Verschelde and F.~Verstraete,
	\textit{Matrix product states for gauge field theories},
	Phys.\ Rev.\ Lett.\  {\bf 113}, 091601 (2014)
	doi:10.1103/PhysRevLett.113.091601
	[arXiv:1312.6654 [hep-lat]].
	
	\bibitem{LGTTN1Ham5} 
	K.~Van Acoleyen, B.~Buyens, J.~Haegeman and F.~Verstraete,
	\textit{Matrix product states for Hamiltonian lattice gauge theories},
	PoS LATTICE {\bf 2014}, 308 (2014)
	doi:10.22323/1.214.0308
	[arXiv:1411.0020 [hep-lat]].
	
	\bibitem{LGTTN1Ham6} 
	H.~Saito, M.~C.~Ba\~nuls, K.~Cichy, J.~I.~Cirac and K.~Jansen,
	\textit{The temperature dependence of the chiral condensate in the Schwinger model with Matrix Product States},
	PoS LATTICE {\bf 2014}, 302 (2014)
	doi:10.22323/1.214.0302
	[arXiv:1412.0596 [hep-lat]].
	
	\bibitem{LGTTN1Ham7} 
	M.~C.~Ba\~nuls, K.~Cichy, J.~I.~Cirac, K.~Jansen and H.~Saito,
	\textit{Thermal evolution of the Schwinger model with Matrix Product Operators},
	Phys.\ Rev.\ D {\bf 92}, no. 3, 034519 (2015)
	doi:10.1103/PhysRevD.92.034519
	[arXiv:1505.00279 [hep-lat]].
	
	\bibitem{LGTTN1Ham8} 
	T.~Pichler, M.~Dalmonte, E.~Rico, P.~Zoller and S.~Montangero,
	\textit{Real-time Dynamics in U(1) Lattice Gauge Theories with Tensor Networks},
	Phys.\ Rev.\ X {\bf 6}, no. 1, 011023 (2016)
	doi:10.1103/PhysRevX.6.011023
	[arXiv:1505.04440 [cond-mat.quant-gas]].
	
	\bibitem{LGTTN1Ham9} 
	S.~K\"uhn, J.~I.~Cirac and M.~C.~Ba\~nuls,
	\textit{Non-Abelian string breaking phenomena with Matrix Product States},
	JHEP {\bf 1507}, 130 (2015)
	doi:10.1007/JHEP07(2015)130
	[arXiv:1505.04441 [hep-lat]].
	
	\bibitem{LGTTN1Ham10} 
	A.~Milsted,
	\textit{Matrix product states and the non-Abelian rotor model},
	Phys.\ Rev.\ D {\bf 93}, no. 8, 085012 (2016)
	doi:10.1103/PhysRevD.93.085012
	[arXiv:1507.06624 [hep-lat]].
	
	\bibitem{LGTTN1Ham11} 
	B.~Buyens, J.~Haegeman, H.~Verschelde, F.~Verstraete and K.~Van Acoleyen,
	\textit{Confinement and string breaking for QED$_2$ in the Hamiltonian picture},
	Phys.\ Rev.\ X {\bf 6}, no. 4, 041040 (2016)
	doi:10.1103/PhysRevX.6.041040
	[arXiv:1509.00246 [hep-lat]].
	
	\bibitem{LGTTN1Ham12} 
	H.~Saito, M.~C.~Ba\~nuls, K.~Cichy, J.~I.~Cirac and K.~Jansen,
	\textit{Thermal evolution of the one-flavour Schwinger model with using Matrix Product States},
	PoS LATTICE {\bf 2015}, 283 (2016)
	doi:10.22323/1.251.0283
	[arXiv:1511.00794 [hep-lat]].
	
	\bibitem{LGTTN1Ham13} 
	B.~Buyens, J.~Haegeman, F.~Verstraete and K.~Van Acoleyen,
	\textit{Tensor networks for gauge field theories},
	PoS LATTICE {\bf 2015}, 280 (2016)
	doi:10.22323/1.234.0375, 10.22323/1.251.0280
	[arXiv:1511.04288 [hep-lat]].
	
	\bibitem{LGTTN1Ham14} 
	M.~C.~Ba\~nuls, K.~Cichy, K.~Jansen and H.~Saito,
	\textit{Chiral condensate in the Schwinger model with Matrix Product Operators},
	Phys.\ Rev.\ D {\bf 93}, no. 9, 094512 (2016)
	doi:10.1103/PhysRevD.93.094512
	[arXiv:1603.05002 [hep-lat]].
	
	\bibitem{LGTTN1Ham15} 
	B.~Buyens, F.~Verstraete and K.~Van Acoleyen,
	\textit{Hamiltonian simulation of the Schwinger model at finite temperature},
	Phys.\ Rev.\ D {\bf 94}, no. 8, 085018 (2016)
	doi:10.1103/PhysRevD.94.085018
	[arXiv:1606.03385 [hep-lat]].
	
	\bibitem{LGTTN1Ham16} 
	P.~Silvi, E.~Rico, M.~Dalmonte, F.~Tschirsich, S.~Montangero,
	\textit{Finite-density phase diagram of a (1+1)-d non-abelian lattice gauge theory with tensor networks},
	Quantum 1, 9 (2017)
	[arXiv:1606.05510 [hep-lat]].
	
	\bibitem{LGTTN1Ham17} 
	M.~C.~Ba\~nuls, K.~Cichy, J.~I.~Cirac, K.~Jansen and S.~K\"uhn,
	\textit{Density Induced Phase Transitions in the Schwinger Model: A Study with Matrix Product States},
	Phys.\ Rev.\ Lett.\  {\bf 118}, no. 7, 071601 (2017)
	doi:10.1103/PhysRevLett.118.071601
	[arXiv:1611.00705 [hep-lat]].
	
	\bibitem{LGTTN1Ham18} 
	B.~Buyens, J.~Haegeman, F.~Hebenstreit, F.~Verstraete and K.~Van Acoleyen,
	\textit{Real-time simulation of the Schwinger effect with Matrix Product States},
	Phys.\ Rev.\ D {\bf 96}, no. 11, 114501 (2017)
	doi:10.1103/PhysRevD.96.114501
	[arXiv:1612.00739 [hep-lat]].
	
	\bibitem{LGTTN1Ham19} 
	B.~Buyens, S.~Montangero, J.~Haegeman, F.~Verstraete and K.~Van Acoleyen,
	\textit{Finite-representation approximation of lattice gauge theories at the continuum limit with tensor networks},
	Phys.\ Rev.\ D {\bf 95}, no. 9, 094509 (2017)
	doi:10.1103/PhysRevD.95.094509
	[arXiv:1702.08838 [hep-lat]].
	
	\bibitem{LGTTN1Ham20} 
	K.~Zapp and R.~Orus,
	Tensor network simulation of QED on infinite lattices: Learning from (1+1) d , and prospects for (2+1) d,
	Phys.\ Rev.\ D {\bf 95}, no. 11, 114508 (2017)
	doi:10.1103/PhysRevD.95.114508
	[arXiv:1704.03015 [hep-lat]].
	
	\bibitem{LGTTN1Ham21}
	E.~Ercolessi, P.~Facchi, G.~Magnifico, S.~Pascazio and F.~V.~Pepe,
	\textit{Phase transitions in $\mathbb{Z}_n$ gauge models: Towards quantum simulations of the Schwinger-Weyl QED}
	Phys. Rev. D 98, 074503 (2018)
	doi:10.1103/PhysRevD.98.074503
	[arXiv:1705.11047 [quant-ph]]
	
	
	\bibitem{LGTTN1Ham22} 
	M.~C.~Ba\~nuls, K.~Cichy, J.~I.~Cirac, K.~Jansen and S.~K\"uhn,
	\textit{Efficient basis formulation for 1+1 dimensional SU(2) lattice gauge theory: Spectral calculations with matrix product states},
	Phys.\ Rev.\ X {\bf 7}, no. 4, 041046 (2017)
	doi:10.1103/PhysRevX.7.041046
	[arXiv:1707.06434 [hep-lat]].
	
	\bibitem{LGTTN1Ham23}
	G.~Magnifico, D. Vodola, E.~Ercolessi, S.~P.~Kumar, M.~M\"uller and A.~Bermudez,
	\textit{Symmetry-protected topological phases in lattice gauge theories: Topological QED$_2$}
	Phys. Rev. D 99, 014503 (2019)
	doi:10.1103/PhysRevD.99.014503
	[arXiv:1804.10568 [cond-mat.quant-gas]]
	
	\bibitem{LGTTN1Ham24} 
	F.~Bruckmann, K.~Jansen and S.~K\"uhn,
	\textit{O(3) nonlinear sigma model in 1+1 dimensions with matrix product states},
	Phys.\ Rev.\ D {\bf 99}, no. 7, 074501 (2019)
	doi:10.1103/PhysRevD.99.074501
	[arXiv:1812.00944 [hep-lat]].
	
	\bibitem{LGTTN1Ham25} 
	P.~Silvi, Y.~Sauer, F.~Tschirsich and S.~Montangero,
	\textit{Tensor Network Simulation of compact one-dimensional lattice Quantum Chromodynamics at finite density},
	arXiv:1901.04403 [quant-ph].
	
	\bibitem{LGTTN1Ham26}
	G.~Magnifico, D. Vodola, E.~Ercolessi, S.~P.~Kumar, M.~M\"uller and A.~Bermudez,
	\textit{$\mathbb{Z}_n$ gauge theories coupled to topological fermions: QED$_2$ with a quantum mechanical $\theta$ angle}
	Phys. Rev. B 100, 115152 (2019)
	doi:10.1103/PhysRevB.100.115152
	[arXiv:1906.07005 [cond-mat.quant-gas]]
	
	\bibitem{LGTTN1Ham27} 
	L.~Funcke, K.~Jansen and S.~K\"uhn,
	\textit{Topological vacuum structure of the Schwinger model with matrix product states},
	arXiv:1908.00551 [hep-lat].
	
	\bibitem{LGTTN1Ham28}
	G.~Magnifico, M.~Dallmonte, P.~Facchi, S.~Pascazio, F.~V.~Pepe, and E.~Ercolessi,
	\textit{Real Time Dynamics and Confinement in the $\mathbb{Z}_n$ Schwinger-Weyl lattice model for 1+1 QED}
	arXiv:1909.04821 [quant-ph]
	
	%Lagrangian 1+1d %%%%%%%%%%%%%%%%%%%%%%%%%%%%%%%%%%%%
	
	\bibitem{LGTTN1Lag1} 
	Y.~Liu, Y.~Meurice, M.~P.~Qin, J.~Unmuth-Yockey, T.~Xiang, Z.~Y.~Xie, J.~F.~Yu and H.~Zou,
	\textit{Exact Blocking Formulas for Spin and Gauge Models},
	Phys.\ Rev.\ D {\bf 88}, 056005 (2013)
	doi:10.1103/PhysRevD.88.056005
	[arXiv:1307.6543 [hep-lat]].
	
	\bibitem{LGTTN1Lag2} 
	Y.~Shimizu and Y.~Kuramashi,
	\textit{Grassmann tensor renormalization group approach to one-flavor lattice Schwinger model},
	Phys.\ Rev.\ D {\bf 90}, no. 1, 014508 (2014)
	doi:10.1103/PhysRevD.90.014508
	[arXiv:1403.0642 [hep-lat]].
		
	\bibitem{LGTTN1Lag3} 
	Y.~Shimizu and Y.~Kuramashi,
	\textit{Critical behavior of the lattice Schwinger model with a topological term at $\theta=\pi$ using the Grassmann tensor renormalization group},
	Phys.\ Rev.\ D {\bf 90}, no. 7, 074503 (2014)
	doi:10.1103/PhysRevD.90.074503
	[arXiv:1408.0897 [hep-lat]].
	
	\bibitem{LGTTN1Lag4} 
	J.~F.~Unmuth-Yockey, Y.~Meurice, J.~Osborn and H.~Zou,
	\textit{Tensor renormalization group study of the 2d O(3) model},
	PoS LATTICE {\bf 2014}, 325 (2014)
	doi:10.22323/1.214.0325
	[arXiv:1411.4213 [hep-lat]].
	
	\bibitem{LGTTN1Lag5} 
	L.~P.~Yang, Y.~Liu, H.~Zou, Z.~Y.~Xie and Y.~Meurice,
	\textit{Fine structure of the entanglement entropy in the O(2) model},
	Phys.\ Rev.\ E {\bf 93}, no. 1, 012138 (2016)
	doi:10.1103/PhysRevE.93.012138
	[arXiv:1507.01471 [cond-mat.stat-mech]].
	
	\bibitem{LGTTN1Lag6} 
	H.~Kawauchi and S.~Takeda,
	\textit{Tensor renormalization group analysis of CP(N-1) model},
	Phys.\ Rev.\ D {\bf 93}, no. 11, 114503 (2016)
	doi:10.1103/PhysRevD.93.114503
	[arXiv:1603.09455 [hep-lat]].
	
	\bibitem{LGTTN1Lag7} 
	A.~Bazavov, Y.~Meurice, S.-W.~Tsai, J.~Unmuth-Yockey, L.~P.~Yang and J.~Zhang,
	\textit{Estimating the central charge from the Rényi entanglement entropy},
	Phys.\ Rev.\ D {\bf 96}, no. 3, 034514 (2017)
	doi:10.1103/PhysRevD.96.034514
	[arXiv:1703.10577 [hep-lat]].
	
	\bibitem{LGTTN1Lag8} 
	Y.~Shimizu and Y.~Kuramashi,
	\textit{Berezinskii-Kosterlitz-Thouless transition in lattice Schwinger model with one flavor of Wilson fermion},
	Phys.\ Rev.\ D {\bf 97}, no. 3, 034502 (2018)
	doi:10.1103/PhysRevD.97.034502
	[arXiv:1712.07808 [hep-lat]].
	
	\bibitem{LGTTN1Lag9} 
	J.~Unmuth-Yockey, J.~Zhang, A.~Bazavov, Y.~Meurice and S.~W.~Tsai,
	\textit{Universal features of the Abelian Polyakov loop in 1+1 dimensions},
	Phys.\ Rev.\ D {\bf 98}, no. 9, 094511 (2018)
	doi:10.1103/PhysRevD.98.094511
	[arXiv:1807.09186 [hep-lat]].
	
	\bibitem{LGTTN1Lag10} 
	A.~Bazavov, S.~Catterall, R.~G.~Jha and J.~Unmuth-Yockey,
	\textit{Tensor renormalization group study of the non-Abelian Higgs model in two dimensions},
	Phys.\ Rev.\ D {\bf 99}, no. 11, 114507 (2019)
	doi:10.1103/PhysRevD.99.114507
	[arXiv:1901.11443 [hep-lat]].
	
	\bibitem{LGTTN1Lag11} 
	M.~Asaduzzaman, S.~Catterall and J.~Unmuth-Yockey,
	\textit{Tensor network formulation of two dimensional gravity},
	arXiv:1905.13061 [hep-lat].
	
	%Hamiltonian 2+1d %%%%%%%%%%%%%%%%%%%%%%%%%%%%%%%%%%%%%
	
	\bibitem{LGTTN2Ham1} 
	M.~Aguado and G.~Vidal,
	\textit{Entanglement renormalization and topological order},
	Phys.\ Rev.\ Lett.\  {\bf 100}, 070404 (2008)
	doi:10.1103/PhysRevLett.100.070404
	[arXiv:0712.0348 [cond-mat.str-el]].
	
	\bibitem{LGTTN2Ham2}
	R K\"onig, BW Reichardt, G Vidal
	\textit{Exact entanglement renormalization for string-net models},
	Physical Review B 79 (19), 195123
	[arXiv:arXiv:0806.4583]
	
	\bibitem{LGTTN2Ham3}
	L.~Tagliacozzo and G.~Vidal,
	\textit{Entanglement renormalization and gauge symmetry},
	Phys.\ Rev.\ B {\bf 83}, 115127 (2011)
	doi:10.1103/PhysRevB.83.115127
	[arXiv:1007.4145]
	
	\bibitem{LGTTN2Ham4} 
	P.~Silvi, E.~Rico, T.~Calarco and S.~Montangero,
	\textit{Lattice Gauge Tensor Networks},
	New J.\ Phys.\  {\bf 16}, no. 10, 103015 (2014)
	doi:10.1088/1367-2630/16/10/103015
	[arXiv:1404.7439 [quant-ph]].	

	\bibitem{LGTTN2Ham5} 
	L.~Tagliacozzo, A.~Celi and M.~Lewenstein,
	\textit{Tensor Networks for Lattice Gauge Theories with continuous groups},
	Phys.\ Rev.\ X {\bf 4}, no. 4, 041024 (2014)
	doi:10.1103/PhysRevX.4.041024
	[arXiv:1405.4811 [cond-mat.str-el]].
	
	\bibitem{LGTTN2Ham6} 
	J.~Haegeman, K.~Van Acoleyen, N.~Schuch, J.~I.~Cirac and F.~Verstraete,
	\textit{Gauging quantum states: from global to local symmetries in many-body systems},
	Phys.\ Rev.\ X {\bf 5}, no. 1, 011024 (2015)
	doi:10.1103/PhysRevX.5.011024
	[arXiv:1407.1025 [quant-ph]].
	
	\bibitem{LGTTN2Ham7} 
	E.~Zohar, M.~Burrello, T.~Wahl and J.~I.~Cirac,
	\textit{Fermionic Projected Entangled Pair States and Local U(1) Gauge Theories},
	Annals Phys.\  {\bf 363}, 385 (2015)
	doi:10.1016/j.aop.2015.10.009
	[arXiv:1507.08837 [quant-ph]].
		
	\bibitem{LGTTN2Ham8} 
	E.~Zohar and M.~Burrello,
	\textit{Building Projected Entangled Pair States with a Local Gauge Symmetry},
	New J.\ Phys.\  {\bf 18}, no. 4, 043008 (2016)
	doi:10.1088/1367-2630/18/4/043008
	[arXiv:1511.08426 [quant-ph]].
	
	\bibitem{LGTTN2Ham9} 
	A.~Milsted and T.~J.~Osborne,
	\textit{Quantum Yang-Mills theory: An overview of a program},
	Phys.\ Rev.\ D {\bf 98}, no. 1, 014505 (2018)
	doi:10.1103/PhysRevD.98.014505
	[arXiv:1604.01979 [quant-ph]].
	
	\bibitem{LGTTN2Ham10} 
	E.~Zohar, T.~B.~Wahl, M.~Burrello and J.~I.~Cirac,
	\textit{Projected Entangled Pair States with non-Abelian gauge symmetries: an SU(2) study},
	Annals Phys.\  {\bf 374}, 84 (2016)
	doi:10.1016/j.aop.2016.08.008
	[arXiv:1607.08115 [quant-ph]].
	
	\bibitem{LGTTN2Ham11} 
	E.~Zohar and J.~I.~Cirac,
	\textit{Combining tensor networks with Monte Carlo methods for lattice gauge theories},
	Phys.\ Rev.\ D {\bf 97}, no. 3, 034510 (2018)
	doi:10.1103/PhysRevD.97.034510
	[arXiv:1710.11013 [quant-ph]].
	
	%Lagrangian 2+1d %%%%%%%%%%%%%%%%%%%%%%%%%%%%%%%%%%
	
	\bibitem{LGTTN2Lag} 
	Y.~Kuramashi and Y.~Yoshimura,
	\textit{Three-dimensional finite temperature Z$_{2}$ gauge theory with tensor network scheme},
	JHEP {\bf 1908}, 023 (2019)
	doi:10.1007/JHEP08(2019)023
	[arXiv:1808.08025 [hep-lat]].
		
	
	%Reviews%%%%%%%%%%%%%%%%%%%%%%%%%%%%%%%%%%%%%%%%%%%%%
	
	\bibitem{LGTTNrev1} 
	P.~Emonts and E.~Zohar,
	Gauss law, Minimal Coupling and Fermionic PEPS for Lattice Gauge Theories,
	arXiv:1807.01294 [quant-ph].
	
	\bibitem{LGTTNrev2}
	M.~C.~Ba\~nuls \textit{et al.},
	\textit{Tensor Networks and their use for Lattice Gauge Theories} in proceedings of the 36th Annual International Symposium on Lattice Field Theory
	PoS(LATTICE2018)022
	doi:10.22323/1.214.0016
	
	\bibitem{LGTTNrev3} 
	M.~Dalmonte and S.~Montangero,
	\textit{Lattice gauge theory simulations in the quantum information era},
	Contemp.\ Phys.\  {\bf 57}, no. 3, 388 (2016)
	doi:10.1080/00107514.2016.1151199
	[arXiv:1602.03776 [cond-mat.quant-gas]].
	
	\bibitem{LGTTNrev4} 
	M.~C.~Ba\~nuls and K.~Cichy
	\textit{Review on novel methods for lattice gauge theories},
	arXiv:1910.00257 [hep-lat].
	
	\bibitem{LGTTNrev5}
	M.~C.~Ba\~nuls, R.~Blatt, J.~Catani, A.~Celi, J.~I.~Cirac, M.~Dalmonte, L.~Fallani, K.~Jansen, M.~Lewenstein, S.~Montangero, C.~A.~Muschik, B.~Reznik, E.~Rico, L.~Tagliacozzo, K.~van~Acoleyen, F.~Verstraete, U.-J.~Wiese, M.~Wingate, J.~Zakrzewski, P.~Zoller
	\textit{Simulating Lattice Gauge Theories within Quantum Technologies},
	arXiv:1911.00003 [hep-lat].
	
%%%%%%%%%%%%%%%%%%%%%%%%%%%%%%%%%%%%%%%%%%%%%%%%%%%%%%%%%%%%%%%%%%%%%%%%%%%%%%%%%%%%%%%%%%%%%%%%%%%%%%%%%%%%%%%%%%%%%%%%%%%%%%
	
	\bibitem{cMERA} 
	J. Haegeman, T. J. Osborne, H. Verschelde and F. Verstraete, 
	\textit{Entanglement Renormalization for Quantum Fields in Real Space}, 
	Phys. Rev. Lett., 110, 100402 (2013),
	arxiv:1102.5524
	
	\bibitem{intercMERA1} 
	J.~S.~Cotler, J.~Molina-Vilaplana and M.~T.~Mueller,
	\textit{A Gaussian Variational Approach to cMERA for Interacting Fields},
	arXiv:1612.02427 [hep-th].
	
	\bibitem{intercMERA2} 
	J.~S.~Cotler, M.~Reza Mohammadi Mozaffar, A.~Mollabashi and A.~Naseh,
	\textit{Entanglement renormalization for weakly interacting fields},
	Phys.\ Rev.\ D {\bf 99}, no. 8, 085005 (2019)
	doi:10.1103/PhysRevD.99.085005
	[arXiv:1806.02835 [hep-th]].
	
	\bibitem{intercMERA3} 
	J.~J.~Fernandez-Melgarejo, J.~Molina-Vilaplana and E.~Torrente-Lujan,
	\textit{Entanglement Renormalization for Interacting Field Theories},
	arXiv:1904.07241 [hep-th].
	
	\bibitem{cMERAholo1}
	A. Mollabashi, M. Naozaki, S. Ryu and T. Takayanagi, 
	\textit{Holographic geometry of cMERA for quantum quenches and finite temperature}, 
	JHEP (2014) 2014: 98, 
	arxiv:1311.6095.
	
	\bibitem{cMERAholo2}
	M. Nozaki, S. Ryu and T. Takayanagi, 
	\textit{Holographic geometry of entanglement renormalization in quantum field theories}, 
	JHEP (2012) 2012: 10, 
	arxiv:1208.3469.
	
	\bibitem{cMERAholo3}
	J. Molina-Vilaplana, 
	\textit{Information geometry of entanglement renormalization for free quantum fields}, 
	JHEP (2015) 2015:2 (mar, 2015), 
	arxiv:1503.07699.
	
	\bibitem{cMERAholo4}
	J. Molina-Vilaplana, 
	\textit{Entanglement renormalization and two dimensional string theory},
	Phys. Lett. B 755 (2016) 421-425, 
	arxiv:1510.09020.
	
	\bibitem{cMERAholo5}
	M. Miyaji, S. Ryu, T. Takayanagi and X. Wen, 
	\textit{Boundary states as holographic duals of trivial spacetimes}, 
	JHEP (2015) 2015: 152, 
	arxiv:1412.6226.
	
	\bibitem{cMERAholo6}
	M. Miyaji, T. Numasawa, N. Shiba, T. Takayanagi, K. Watanabe,
	\textit{cMERA as Surface/State Correspondence in AdS/CFT},
	Phys. Rev. Lett. 115, 171602 (2015),
	arXiv:1506.01353.
	
	\bibitem{cMERAholo7}
	M. Miyaji and T. Takayanagi, 
	\textit{Surface/state correspondence as a generalized holography},
	Progress of Theoretical and Experimental Physics 2015 (mar, 2015) , arxiv:1503.03542.
	
	\bibitem{cMERAholo8}
	X. Wen, G. Y. Cho, P. L. S. Lopes, Y. Gu, X. L. Qi and S. Ryu, 
	\textit{Holographic entanglement renormalization of topological insulators}, 
	Phys. Rev. B 94, 075124 (2016), 
	arxiv:1605.07199.
	
	\bibitem{cMERAholo9}
	J. R. Fliss, R. G. Leigh and O. Parrikar, 
	\textit{Unitary Networks from the Exact Renormalization of Wave Functionals}, 
	Phys. Rev. D 95, 126001 (2017), 
	arxiv:1609.03493.
		
	\bibitem{Zou2019}
	Y.~Zou, M.~Ganahl and G.~Vidal,
	\textit{Magic entanglement renormalization for quantum fields},
	arXiv:1906.04218 [cond-mat.str-el].
	
	\bibitem{Adrian}
	A. Franco-Rubio, G. Vidal,
	\textit{Entanglement and correlations in the continuous multi-scale entanglement renormalization ansatz},
	JHEP 2017 (12), 129, 
	arxiv:1706.02841.	
	
	\bibitem{Qi}
	Q. Hu, G. Vidal, 
	\textit{Spacetime symmetries and conformal data in the continuous multi-scale entanglement renormalization ansatz}
	Phys. Rev. Lett. 119, 010603 (2017),
	arxiv:1703.04798
	
	\bibitem{Ganahl2019}
	Y.~Zou, M.~Ganahl and G.~Vidal,
	\textit{Interacting cMERA},
	\textit{in preparation}.

	
	\bibitem{cMPS1} 
	F.~Verstraete and J.~I.~Cirac,
	\textit{Continuous Matrix Product States for Quantum Fields},
	Phys.\ Rev.\ Lett.\  {104}, 190405 (2010)
	doi:10.1103/PhysRevLett.104.190405
	[arXiv:1002.1824 [cond-mat.str-el]].
	
	\bibitem{cMPS2} 
	J. Haegeman, J. I. Cirac, T. J. Osborne, H. Verschelde,
	and F. Verstraete, 
	\textit{Applying the variational principle to (1+1)-dimensional quantum field theories},
	Phys. Rev. Lett. 105, 251601 (2010), 
	arXiv:1006.2409.
	
	\bibitem{cMPS3}
	J. Haegeman, J. I. Cirac, T. J. Osborne, I. Pizorn, H. Verschelde, F. Verstraete,
	\textit{Time-dependent variational principle for quantum lattices}
	Phys. Rev. Lett. 107, 070601 (2011),
	arXiv:1103.0936.
	
	\bibitem{cMPS4}
	D. Draxler, J. Haegeman, T. J. Osborne, V. Stojevic, L. Vanderstraeten, and F. Verstraete, 
	\textit{Particles, holes and solitons: a matrix product state approach},
	Phys. Rev. Lett. 111, 020402 (2013),
	arXiv:1212.1114.
	
	\bibitem{cMPS5}
	F. Quijandr\'ia, J. J. Garc\'ia-Ripoll, and D. Zueco, 
	\textit{Continuous matrix product states for coupled fields: Application to Luttinger liquids and quantum simulators}
	Phys. Rev. B 90, 235142 (2014),
	arXiv:1409.4709.
	
	\bibitem{cMPS6}
	S. S. Chung, K. Sun, and C. J. Bolech, 
	\textit{Matrix product ansatz for Fermi fields in one dimension}
	Phys. Rev. B 91, 121108(R) (2015),
	arXiv:1501.00228
	
	\bibitem{cMPS7} 
	F. Quijandr\'ia and D. Zueco, 
	\textit{Continuous-matrix-product-state solution for the mixing-demixing transition in one-dimensional quantum fields}
	Phys. Rev. A 92, 043629 (2015),
	arXiv:1507.03613
	
	\bibitem{cMPS8} J. Haegeman, D. Draxler, V. Stojevic, J. I. Cirac,
	T. J. Osborne, and F. Verstraete, 
	\textit{Quantum Gross-Pitaevskii Equation},
	SciPost Phys. 3, 006 (2017),
	arXiv:1501.06575
	
	\bibitem{cMPS9} 
	J. Rinc\'on, M. Ganahl, and G. Vidal, 
	\textit{Lieb-Liniger model with exponentially decaying interactions: A continuous matrix product state study},
	Phys. Rev. B 92, 115107 (2015),
	arXiv:1508.04779. 
	
	\bibitem{cMPS10}
	S. S. Chung, K. Sun, and C. J. Bolech, 
	\textit{Matrix product ansatz for Fermi fields in one dimension},
	Phys. Rev. B 91, 121108 (2015),
	arXiv:1501.00228.
	
	\bibitem{cMPS11}
	D. Draxler, J. Haegeman, F. Verstraete, and M. Rizzi,
	\textit{Continuous matrix product states with periodic boundary conditions and an application to atomtronics},
	Phys. Rev. B 95, 045145 (2017),
	arXiv:1609.09704.
	
	\bibitem{cMPS12} 
	M. Ganahl, J. Rinc\'on, and G. Vidal, 
	\textit{Continuous Matrix Product States for Quantum Fields: An Energy Minimization Algorithm},
	Phys. Rev. Lett. 118, 220402 (2017),
	arXiv:1611.03779.
	
	\bibitem{cMPS13}
	M. Ganahl, 
	\textit{Continuous Matrix Product States for Inhomogeneous Quantum Field Theories: a Basis-Spline Approach},
	arXiv:1712.01260.
	
	\bibitem{cMPS14}
	M. Ganahl, G. Vidal,
	Continuous matrix product states for non-relativistic quantum fields:
	a lattice algorithm for inhomogeneous systems
	Phys. Rev. B 98, 195105 (2018),
	arXiv:1801.02219.
\end{thebibliography}
\end{document}